%% file: ms.tex
\documentclass[journal,letter,10pt]{IEEEtran}

\usepackage{amsmath}
\usepackage[amssymb]{SIunits}
\usepackage{xcolor}
\usepackage{bm} %Bold Math, needed for \bm
\usepackage{amsfonts}
\usepackage{mathtools}
\usepackage{graphicx}
\usepackage{flushend}
\usepackage{caption}
\usepackage{cespvect}
\usepackage{import}
\usepackage{multirow}
\usepackage{cite}
\usepackage[export]{adjustbox}
\usepackage{breqn}

\graphicspath{{./figures/}}

\newcommand{\revision}[1]{\textcolor{black}{#1}}

\newcommand{\eq}[1]{Eq.~\eqref{#1}}
\newcommand{\fig}[1]{Fig.~\ref{#1}}

\newcommand{\tab}[1]{Tab.~\ref{#1}}
\newcommand{\secref}[1]{Section~\ref{#1}}

\newcommand{\myparstart}[1]{\noindent \textbf{#1}}

\providecommand{\newoperator}[3]{%
  \newcommand*{#1}{\mathop{#2}#3}}
\newoperator{\argmax}{\mathrm{argmax}}{\nolimits}

\newcommand{\prop}[1]{Proposition~\ref{#1}}

\newcommand{\DGSet}{{\mathcal{G}}} 	%DG Set
\newcommand{\LSet}{{\mathcal{L}}}	%Loads Set
\newcommand{\PSet}{{\mathcal{P}}} 	%Unitary Price
\newcommand{\HSet}{{\mathcal{H}}} 	%Quantity to Sell
\newcommand{\DSet}{{\mathcal{D}}} 	%Quantity to Buy
\newcommand{\SSet}{{\mathcal{S}}} 	%Unitary Discount

\newcommand{\Pmat}{{\bm{P}}}
\newcommand{\Hmat}{{\bm{H}}}
\newcommand{\Dmat}{{\bm{D}}}
\newcommand{\Smat}{{\bm{S}}}

\newcommand{\Pel}[2]{{p_{#1,#2}}}
\newcommand{\Hel}[2]{{h_{#1,#2}}}
\newcommand{\Del}[2]{{d_{#1,#2}}}
\newcommand{\Sel}[2]{{s_{#1,#2}}}

\newcommand{\Prow}[1]{{\bm{p}_{#1,\cdot}}}
\newcommand{\Hrow}[1]{{\bm{h}_{#1,\cdot}}}
\newcommand{\Drow}[1]{{\bm{d}_{#1,\cdot}}}
\newcommand{\Srow}[1]{{\bm{s}_{#1,\cdot}}}

\newcommand{\Pcol}[1]{{\bm{p]}_{\cdot,#1}}}
\newcommand{\Pcolprime}[1]{{\bm{p}^{\bm{\prime}}_{\cdot,#1}}}
\newcommand{\Hcol}[1]{{\bm{h}_{\cdot,#1}}}
\newcommand{\Dcol}[1]{{\bm{d}_{\cdot,#1}}}

\newcommand{\HrowT}[1]{{\bm{\tilde{H}}_{#1,\cdot}}}
\newcommand{\HcolT}[1]{{\bm{\tilde{H}}_{\cdot,#1}}}
\newcommand{\DrowT}[1]{{\bm{\tilde{D}}_{#1,\cdot}}}
\newcommand{\DcolT}[1]{{\bm{\tilde{D}}_{\cdot,#1}}}

\newcommand{\DrowD}[1]{{\bm{d}^{\bm{\diamond}}_{#1,\cdot}}}

\newcommand{\HmatO}{{\bm{H^{*}}}}
\newcommand{\DmatO}{{\bm{D^{*}}}}

\newcommand{\PelO}[2]{{p^{*}_{#1,#2}}}
\newcommand{\HelO}[2]{{h^{*}_{#1,#2}}}
\newcommand{\DelO}[2]{{d^{*}_{#1,#2}}}
\newcommand{\SelO}[2]{{s^{*}_{#1,#2}}}

\newcommand{\HrowO}[1]{{\bm{h}^{\bm{*}}_{#1,\cdot}}}
\newcommand{\DrowO}[1]{{\bm{d}^{\bm{*}}_{#1,\cdot}}}

\newcommand{\PelPrime}[2]{{p_{#1,#2}^\prime}}

\newcommand{\UiG}{U_{i}^{\DGSet}(\Prow{i},\Hrow{i})}
\newcommand{\UiL}{U_{i}^{\LSet}(\Pcol{i},\Drow{i},\Srow{i})}
\newcommand{\UiLprime}{U_{i}^{\LSet}(\Pcolprime{i},\Drow{i},\Srow{i})}
\newcommand{\UiPCC}{U_{i}^{\text{PCC}}(\Drow{i})}

\newcommand{\cvxUiG}{{U'_{i}}^{\DGSet}(\Prow{i},\Hrow{i})}
\newcommand{\cvxUiL}{{U'_{i}}^{\LSet}(\Pcolprime{i},\Drow{i},\Srow{i})}

\newtheorem{proposition}{Proposition}

\renewcommand{\baselinestretch}{0.98}

\title{Joint Optimal Pricing and Electrical Efficiency Enforcement for Rational Agents in Micro Grids}
\author{Riccardo Bonetto$^{\star\dag}$, Michele Rossi$^{\dag}$, Stefano Tomasin$^{\dag}$, Carlo Fischione$^\ddag$
\thanks{$^\star$Corresponding author. $^\dag$Department of information Engineering, University of Padova, via Gradenigo 6/b, 35131, Padova, Italy. $^\ddag$Electrical Engineering and ACCESS Linnaeus Center, KTH Royal Institute of Technology, Osquldas V\"{a}g 10, 10044, Stockholm, Sweden. This work has been supported by the University of Padova through the Junior Research Grant ENGINE.}}

\begin{document}

\maketitle

\begin{abstract}
\input{abstract}
\end{abstract}

\section{Introduction}\label{sec:introduction}
\input{introduction}

\section{System Model and Use Cases}\label{sec:scenario}
\input{scenario}

\section{Notation and Multi-Objective Optimization Problem}\label{sec:problem}
\input{notation}
\input{moo_problem}

\section{Simulation Setup}\label{sec:sim_setup}
\input{sim_setup}

\section{Results}\label{sec:results}
\input{results}

\section{Conclusions}\label{sec:conclusions}
\input{conclusions}

\bibliographystyle{IEEEtran}
\bibliography{bibliography}
\flushend

\end{document}

%% file: abstract.tex
%!TEX root = main.tex
In electrical distribution grids, the constantly increasing number of power generation devices based on renewables demands a transition from a centralized to a distributed generation paradigm. In fact, power injection from Distributed Energy Resources (DERs) can be selectively controlled to achieve other objectives beyond supporting loads, such as the minimization of the power losses along the distribution lines and the subsequent increase of the grid hosting capacity. However, these technical achievements are only possible if alongside electrical optimization schemes, a suitable market model is set up to promote cooperation from the end users. In contrast with the existing literature, where energy trading and electrical optimization of the grid are often treated separately \revision{or the trading strategy is tailored to a specific electrical optimization objective}, in this work we consider their joint optimization. \revision{We also allow for a modular approach, where the market model can support any smart grid optimization goal.} Specifically, we present a multi-objective optimization problem accounting for energy trading, where: 1) DERs try to maximize their profit, resulting from selling their surplus energy, 2) the loads try to minimize their expense, and 3) the main power supplier aims at maximizing the electrical grid efficiency through a suitable discount policy.
This optimization problem is proved to be non convex, and an equivalent convex formulation is derived. \revision{Centralized solutions are discussed first, and are subsequently distributed transforming the optimization problem into an equivalent one that can be efficiently solved through the alternating direction method of multipliers.} Numerical results to demonstrate the effectiveness of the so obtained optimal policies are then presented, showing the proposed model results in economic benefits for all the users (generators and loads) and in an increased electrical efficiency for the grid.

%% file: introduction.tex
%!TEX root = main.tex
In traditional power grids, two main challenges are emerging: increasing power demand, and uncoordinated injection of electrical power from distributed generators. On the one hand, the constantly increasing power demand calls for radical changes on how the energy is generated and delivered to the final users. On the other hand, the uncoordinated injection of electrical power from distributed generators based on renewables~\cite{EIA-outlook, IEA-stat} tends to destabilize the power network, possibly leading to outages.

To address these problems, recent work has shown that Distributed Energy Resources (DERs) can be effectively used to boost the grid efficiency~\cite{IEEE-book-2013,AKAGI1,BAYOD,SMART-TRENDS,SURVEY-PE-SYSTEM}. This research work has resulted in the proposal of several grid optimization techniques~\cite{PE-DISPERSED, LOAD-UNBALANCE, MICRO-STOR-MAN}, each exploiting some existing communication infrastructure and relying on online smart metering procedures~\cite{GRID-IMP-EST}. A common trait of these techniques is that a coordinated and intelligent control of the distributed generation capabilities (from renewables) holds the potential of enhancing the electrical grid performance, ameliorating the aforementioned  problems and, at the same time, increasing the grid hosting capacity. 

In this paper, we target residential micro grids where some of the end-users behave as DERs, through the exploitation of renewable energy such as solar, wind, biomass, geothermal, etc. In these micro grids, each DER is normally equipped with an energy storage device (i.e., a battery) and it is assumed to fulfill its own power demand. In addition, during each network cycle, DERs can independently decide to either sell part of the stored energy to the main power supplier, which is addressed as the Point of Common Coupling (PCC), or directly to some selected end-users (loads). Without any further regulation, DERs would sell all their energy to the agents ensuring the highest revenue. However, this could lead to inefficient operating points for the grid (e.g., high distribution power losses or instability problems). Control techniques for DERs~\cite{CENTRALIZED,CENTRALIZEDvsDISTRIBUTED,LC, SurroundControl, DORPF} can prevent this, by significantly reducing distribution power losses, relieving the PCC from some of the power load and granting stability. According to these strategies, after local load satisfaction, end-users fine tune their energy injection into the electricity grid so as to reduce the distribution power losses and the total power demand from the mains. %The amount of energy to inject is computed through centralized or distributed control approaches~\cite{CENTRALIZEDvsDISTRIBUTED}. 

Nevertheless, previous control studies for electrical power grids ignore that, {\textit{in real-world scenarios, DERs energy injection's attitude depends on economic advantages}}. Thus, previous approaches may not be viable in practice if not paired with suitable market rules.
New market models for the smart grid have been studied so far in terms of demand-response control and dynamic pricing strategies. Some work addressed the case where a single energy provider determines the best real time pricing policy, maximizing its own economic benefit~\cite{pricing-1,pricing-3} or a specific quality of service function accounting for the main supplier revenue and the aggregated end-users experience \cite{pricing-4,pricing-7}. Other papers exploit dynamic pricing policies to control the power demand from end-users, thus reducing the chance of instability events as, for example, power outages~\cite{pricing-2,pricing-5}. \revision{Moreover, effort has been devoted to the definition of pricing models that enforce the efficiency of specific electrical optimization techniques, see, for example, \cite{paper-rev-1,paper-rev-2,paper-rev-3}}. 
\revision{Here, \textit{we recognize that real users are expected to change their behavior and positively contribute to the grid optimization if this leads to economic benefits} (i.e., a monetary income). Moreover, we note that several electrical optimization techniques may already be deployed in the same micro grid, such as, peak shaving~\cite{ICIT} or power loss minimization~\cite{tii}.}

Thus, we propose an optimization framework that jointly accounts for economic rewards (i.e., lowering the energy consumers expenses and guaranteeing higher profit to the DERs) and for the execution of a selected electrical optimization technique (so to increase the energy efficiency of the power grid and assure its stability). \revision{The proposed framework defines a market model that can be optimized to support any combination of electrical optimization techniques, as long as they determine the amount of power that each DER has to inject at any given time. Also, as we discuss below, the proposed model does not require that preexisting contract terms and conditions for the electric supply are renegotiated. Hence, the approach can be readily deployed with minimal modifications to existing grids/regulations.} Although we are aware that in current markets DERs sell their surplus energy to the PCC (i.e., direct selling to distributed users is not permitted), we assume that energy trading among end-users will be allowed in the future and we disregard regulatory restrictions. In fact, current research trends are promoting distributed architectures where end users trade energy in a peer-to-peer fashion~\cite{PowerToPeople-2014}, as we do here.

The proposed market scenario is naturally formulated as a multi-objective optimization problem. Each grid user (i.e., loads, DERs, and the PCC) is assumed to act as a {\textit{rational agent}} and, in turn, it always tries to maximize its own benefit. Hence, each DER maximizes its own profit, each load minimizes its own expense and the PCC aims at assuring the grid's electrical efficiency. Energy can be traded directly among end-users (i.e., DERs and loads) or between end-users and the PCC. \revision{In this paper, after formally defining the optimization problem, its centralized solution is presented. In this case, the PCC acts as a central controller and regulator for the grid. This solution provides a complete description of the \revision{Pareto-}optimal \revision{(P-optimal)} trading strategy (i.e., energy prices, and energy allocation matrix) for each grid agent. In this centralized case, the PCC solves the problem and then distributes the \revision{P-}optimal parameters to the network agents. Since the solution is guaranteed to provide economic benefit to all the DERs and loads, it is in their best interest to adopt the \revision{P-}optimal trading strategy. In particular, each DER receives from the PCC the energy prices that such a DER should apply and the amount of energy it should sell to each load in the same grid.} The loads can then decide whether to buy energy from the DERs (according to the proposed prices) or from the PCC (according to a fixed common price). The PCC enforces the grid electrical efficiency by applying a discount policy to the price paid by the loads when buying from DERs. \revision{After characterizing the centralized solution, we present a distributed formulation of the problem. This is achieved through a transformation of the original problem into an equivalent one, which is shaped as a general form consensus with regularization~\cite{BOYD-ADMM}. This new problem can be efficiently solved in a decentralized manner using the alternating direction method of multipliers~\cite{BOYD-ADMM}.}

We seek for a \revision{P-}optimal policy that provides the best trading strategy (in terms of economic benefit) for each end-user, while also driving the system toward the best electrical condition (according to a selected grid optimization technique). 
We remark that the proposed model is transparent to the chosen grid optimization strategy. In fact, any electrical optimization technique can be plugged into our framework as long as it provides an optimal power allocation for the nodes. Hence the applicability of our model does not reduce to a single scenario and it does not exclude future improvements in terms of electrical optimization. Moreover, we allow the PCC (i.e., the electrical utility) to set the importance of each performance objective, i.e., end-user revenue {\textit{vs}} grid electrical efficiency. This is achieved by means of a maximum discount factor that limits the individual discount that can be applied in the energy trading between DERs and loads. 

%Although our solution is centralized, we deem the proposed optimization framework valuable to the design of future distributed control strategies in the considered multi-agent system.

%Our optimization must be performed at every network cycle in order to obtain the greatest benefit from it. However, the PCC can also set its own prices for buying and selling energy, so that our model can account for dynamic pricing and other long-term demand-response optimization techniques.

The rest of this paper is structured as follows. \secref{sec:scenario} introduces the considered electrical / market scenarios and \revision{two use cases for the proposed optimization framework.} \secref{sec:problem} presents \revision{the mathematical notation for the market model, the associated multi-objective optimization problem, its discussion and solution.} In this section, we \revision{first} show that the considered optimization problem is non-convex. Thus, a bijective transformation yielding a convex version of the original problem is found and the solution of the new convex problem is \revision{assessed}. \revision{Finally, a decentralized approach based on ADMM is proposed.}
In \secref{sec:sim_setup}, the electrical grid topology and the parameters used to obtain the numerical results are \revision{given}.
In \secref{sec:results}, the numerical results obtained through the setup of \secref{sec:sim_setup} are shown.
Finally, in \secref{sec:conclusions}, we draw the conclusions of our work and discuss the validity of the proposed model.

%% file: scenario.tex
%!TEX root = main.tex
In this section, the electrical, communication and market scenarios are introduced. \revision{Some relevant use cases are described, showing how two electrical optimization techniques from the literature can be plugged into the proposed optimization framework. In particular, we define the communication requirements and infrastructure needed to support the proposed model and discuss the  interactions among the involved (rational) agents.}

\subsection{Electrical Scenario}
\label{ssec:cap_5_electrical_scenario}

\begin{figure}[t]
\begin{center}
\def\svgwidth{0.45\textwidth}
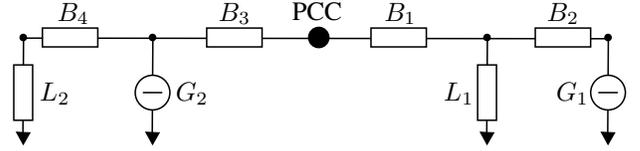
\end{center}
\caption{Electrical network example, where $B_i,\ i=1,\dots,4$ are the electrical distribution lines, $L_j,\ j=1,2$ are the loads, $G_k,\ k=1,2$ are the DERs, and the PCC is the Point of Common Coupling.}
\label{figure:cap_5_electrical_network}
\end{figure}

We consider a steady-state\footnote{In steady-state, the network has reached equilibrium and transient phenomena are no longer relevant.} low voltage power micro grid. For \revision{ease of computation}, and without loss of generality, the considered grid is modeled as a directed tree. The root of the tree represents the Point of Common Coupling (PCC) and the other nodes represent loads and Distributed Energy Resources (DERs). Loads are either \revision{modeled} by constant complex impedances or by constant current sources, the PCC is modeled as a voltage generator setting the voltage \revision{and phase} references for the entire grid, while DERs are modeled either as power or current generators. \revision{Here, we assume that the PCC is always able to supply the grid with the needed power. Hence, no power outages or voltage instabilities (i.e., overvoltages and voltage sags due to the DERs operations) can occur in the considered electrical setup.} This model has been widely considered in the literature, and in particular for power loss minimization algorithms~\cite{TokenRing,SurroundControl,dyngridmap, DORPF,DORPFConvergence}.

\fig{figure:cap_5_electrical_network} \revision{shows} an example power grid is shown. DER $i=1,2$ and load $j=1,2$ are respectively denoted by $G_i$ and $L_j$. Distribution lines are assumed to have a constant section~\cite{SurroundControl, DORPF}, and hence each line has a constant impedance per unit length. The length of the $z$-th distribution line is denoted by $B_z$. Each DER is equipped with a finite-size energy storage device (e.g., rechargeable battery). The size of the energy storage devices determines the total amount of available power. Moreover, each DER is assumed to be feeding an associated load and to have the capability of injecting part of its energy surplus into the grid. The surplus power that DER $G_i$ can inject into the grid is denoted by $E_i$. For the sake of terminology, the quantity $E_i$ will be referred to as $G_i$'s surplus energy. \revision{The amount of energy $E_i$ that each DER wishes to inject into the grid is not regulated by a central authority, but it depends on his local decision. This decision, in turn, depends on the specific energy storage policy that each DER implements, as, for example, the minimization of the probability of not being able to feed its associated load within a given time horizon.} In this paper, it is assumed, without loss of generality, that $E_i\geq0\ \forall \, i$. Each load $L_j$ is assumed to have a non-negative power demand, which is denoted by $D_j \geq 0$. These assumptions are common in the literature, see for example~\cite{ICIT,DORPF,LC,TokenRing}.

%It is worth noting that the proposed model is not tailored to a specific (electrical) grid optimization strategy as long as it deals with the selection of the amount of power that each DER must inject into the grid to boost its electrical efficiency.

%is transparent with respect to the selected (electrical) grid optimization strategy as long as it deals with the selection of the amount of power that each DER must inject into the grid to boost its electrical efficiency.

\subsection{Communication Scenario}
\label{ssec:cap_5_comm_scenario}

Each node (i.e., loads, DERs and the PCC) in the grid is equipped with a transceiver, whose communication performance  depends on the requirements of the selected electrical optimization technique. These details are however neither considered here nor fundamental to the solution of the presented optimization problem. In fact, our optimization framework is independent of the specific  communication technology and infrastructure, as long as these allow a timely bi-directional communication between each pair of nodes.

\subsection{Market Scenario}
\label{ssec:cap_5_market_scenario}

We propose a market scenario where each DER can either sell its surplus power to the PCC or directly to the loads. The monetary revenue that each DER obtains by selling (part of) its energy to the PCC is determined by a PCC-imposed unitary {\it buying price}. The monetary revenue that each DER obtains by selling its energy directly to a specific load is determined by a DER-imposed unitary {\it selling price}. Each DER can independently set a different selling price for each load, i.e., this price is not controlled by the PCC. Also, each load can fulfill its power demand by buying the needed power from the PCC or directly from the DERs. DERs and loads are assumed to behave as {\it rational agents}. That is, each DER will sell its power to the agents (PCC and loads) ensuring the highest revenue, while each load will buy the power it needs from the agents (PCC and DERs) ensuring the lowest expense. Note that this trading model is consistent with the expected evolution of the smart grid market~\cite{PowerToPeople-2014}.

\begin{figure}[tbp]
\begin{center}
\unitlength=1mm
\begin{picture}(82,58)(0,0)
%\put(0,0){\framebox(82,58){}} %framebox to see actual figure size - very useful (uncomment) when creating the composition
%\w(80,92)[12]{\textbf{weekends:}}
\put(0,0){\includegraphics[width=0.9\columnwidth]{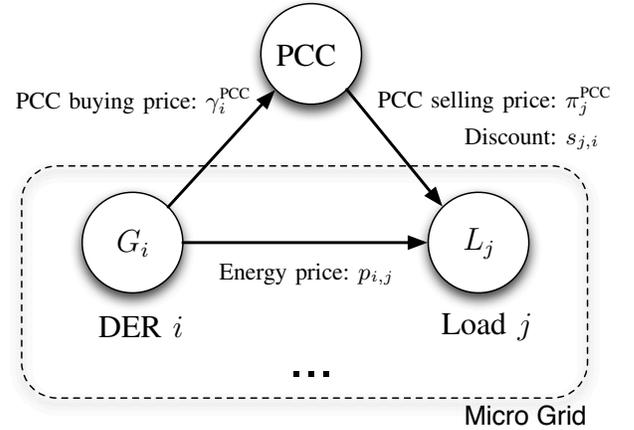}}
\w(17,44)[12]{PCC buying price: $\gamma_{i}^{\text{PCC}}$}
\w(65,44)[12]{PCC selling price: $\pi_{j}^{\text{PCC}}$}
\w(70,39)[12]{Discount: $\Sel{j}{i}$}
\w(40,21)[12]{Energy price: $\Pel{i}{j}$}
\w(40,50)[10]{PCC}
\w(17,25)[10]{$G_i$}
\w(18,14)[10]{DER $i$}
\w(64,14)[10]{Load $j$}
\w(63,25)[10]{$L_j$}
%\w(126,30)[12]{$w_3$}
%\w(135,30)[12]{$w_4$}
\end{picture}
\end{center}
\caption{Market scenario example. Here, the PCC can buy energy from DER $G_i$ paying unitary price $\gamma_i^{\text{PCC}}$, and it can sell energy to load $L_j$ for a unitary price $\pi_j^{\text{PCC}}$. Moreover, $L_j$ can buy energy from $G_i$ paying a discounted unitary price $\Pel{i}{j}-\Sel{j}{i}$, where the discount $\Sel{j}{i}$ is imposed by the PCC.}
\label{figure:cap_5_market_scenario}
\end{figure}

%\begin{figure}[h!]
%\begin{center}
%\def\svgwidth{0.4\textwidth}
%\input{./figures/scenario-opt.eps_tex}
%\end{center}
%\caption{Market scenario example.}
%\label{figure:cap_5_market_scenario}
%\end{figure}

\fig{figure:cap_5_market_scenario} shows an example for the considered market scenario. On the one hand, for each DER $G_i$, the PCC determines the unitary price $\gamma_{i}^{\text{PCC}}$. This is the unitary price that the PCC pays when buying power from DER $G_i$. On the other hand, for each load $L_j$ the PCC determines the unitary price $\pi_{j}^{\text{PCC}}$. This is the unitary price that load $L_j$ pays when buying power from the PCC. \revision{The prices $\gamma_i^{\text{PCC}}$ and $\pi_j^{\text{PCC}}$ do not depend on the optimization process. They are imposed by the PCC according to existing energy trading contracts. Each network agent may have different contract terms and conditions for buying and selling energy from and to the PCC. These conditions set the baseline for the proposed optimization to determine the new trading strategy. With respect to the current practice, where each DER sells $E_i$ energy to the PCC for $\gamma_i^{\text{PCC}}$ revenue and each load buys $D_j$ energy from the PCC spending $\pi_j^{\text{PCC}}D_j$, the new trading strategy results in economic benefits for the network agents, while also enforcing the  electric optimization policy already in place. In the proposed market model,} each DER $G_i$ proposes a unitary price $\Pel{i}{j}$ to each load $L_j$. The unitary price $\Pel{i}{j}$ determines the monetary revenue that $G_i$ obtains when selling power to $L_j$. In order to move the grid electrical state toward the optimal solution (dictated by the selected electrical grid optimization technique), the PCC can apply a discount to the unitary prices $\Pel{i}{j}$ proposed by the DERs to the loads. The discount proposed by the PCC to $L_j$ when buying from $G_i$ is denoted by $\Sel{j}{i}$ and, in turn, the unitary price that $L_j$ pays to $G_i$ is $\Pel{i}{j}-\Sel{j}{i}$. Similar interaction models have been previously used in the scientific literature see, e.g.,~\cite{Pricing-2015}. \revision{We remark that, since DERs and loads are assumed to be {\it rational} agents, any energy trading solution resulting in higher revenues for the DERs and in lower expenses for the loads with respect to the current practice will be embraced. Here, we propose an optimization problem whose solution guarantees better trading conditions for DERs and loads with respect to the current energy trading paradigm and, at the same time, enforces the operation of an electrical optimization technique that is possibly already in place. With respect to the current energy market, the only regulatory act to enable the proposed optimized market is to allow direct energy trading between the grid agents. As noted before, this scenario is expected to become a solid reality in the near future.}

To optimize the electrical status of the power network, we allow the PCC to influence the amount of energy that the DERs inject into the grid. In particular, PCC can prevent DERs from injecting energy by buying it. In this case, the energy bought by the PCC is for example stored by the DERs into their batteries, and could be made available for future use. 

%In the proposed model, the amount of power that $G_i$ sells to the PCC is not injected into the grid. It is, instead, stored into $G_i$'s battery assuming that, at any time, the PCC can claim its ownership and, in turn, force $G_i$ to inject a fraction of that power into the grid. This assumption is needed so that the PCC can buy power from the DERs while, at the same time, enforcing the electrical grid's efficiency. In fact, if the DERs would inject into the grid the power they sell to the PCC, then all the available power would always be injected into the grid, making any optimization of its electrical status impossible.

%All the assumptions and motivation of this section are coherent with standard assumptions and motivation in the literature (CITA). In the following section, the needed mathematical notation will be introduced.

\subsection{Use Cases}
\revision{\myparstart{i) }In the first use case, we consider the current based surround control (CBSC) proposed in~\cite{SurroundControl}. When this technique is implemented, the electrical network is divided into clusters. Each cluster is made of a pair of DERs connected such as on the path between them there are only loads. Within each cluster, the DERs cooperate to feed the loads on the path connecting them. The framework proposed in this paper can be utilized to enforce CBSC by defining market rules that guarantee economic advantages for the loads and the DERs. First, the DERs in each cluster have to solve an electrical optimization problem. To do that, the DERs collect the amount of power needed by the loads. Then, they determine the optimal amount of power that they should inject into the grid to minimize the distribution losses as described in~\cite{SurroundControl}. Once this process is complete, the DERs send to the PCC the information they obtained (i.e., the optimal amount of power that they should inject and the power needed by the loads) together with the maximum amount of power they are willing to sell (i.e., to inject). The PCC collects this information and, by solving the optimization problem defined in this paper determines the prices $p_{i,j}$ and the discounts $s_{j,i}$ of~\fig{figure:cap_5_market_scenario} which guarantee that, when acting as rational agents, DERs and loads experience economic benefits while minimizing the grid power losses. This process can be carried out according to different time scales (i.e., real time, on a minute basis, or on an hourly basis), moreover the electrical optimization time scale and the market one can be decoupled. For example, the electrical optimization can operate in real time, while the market one can operate on an hourly basis. In this case, DERs and loads must send to the PCC hourly generation and consumption data. This can be done in a day ahead fashion by forecasting the power generation and consumption of the next hour in each cluster. This information is then sent to the PCC, who solves the electrical optimization problem assuming a constant power consumption and generation during the next hour and determines the prices $p_{i,j}$ and the discounts $s_{j,i}$ to be applied accordingly. Once this is done, the electrical optimization is performed in real time, but the energy trading strategy remain the same until the next update of market quantities (prices).}

\revision{\myparstart{ii) } The second use case deals with the peak shaving technique in~\cite{ICIT}. This optimization requires that the DERs send to the PCC a day ahead forecast of the generated power. Once the PCC receives this information, it performs a day ahead forecast of the grid power consumption and, based on these two predictions, it computes two parameters that are broadcast to the DERs. Once they receive these parameters, the DERs determine the amount of active and reactive power that they will inject into the grid. Differently from CBSC, in this case only the aggregated power consumption from the loads is required. Nevertheless, the computation of a suitable trading strategy, through the optimization that we propose in this paper, requires one to know the power consumption of {\it each} load. According to the growing diffusion of home deployed smart meters~\cite{SMART-METERS}, this fine-grained information is easily gathered and is expected to be available in nearly all future power grids. With this information, it is possible to perform the market optimization proposed here. To do that, the PCC determines the optimal amount of power that the DERs have to generate to prevent power consumption peaks. Moreover, it determines the amount of power that the DERs have to sell to the loads to fulfill their power demand. For example, consider~\fig{figure:cap_5_electrical_network} and assume that the electrical optimization process dictates that DERs $G_1$ and $G_2$ inject $E_1$ and $E_2$ power, respectively. Moreover, assume that loads $L_1$ and $L_2$ will need $D_1$ and $D_2$ power, respectively. Then, a possible power allocation scheme could be:
\begin{itemize} 
\item $\min(E_1,D_1)$ power is sold by $G_1$ to $D_1$;
\item $\min(E_2,D_2)$ power is sold by $G_2$ to $D_2$;
\item the remaining available (needed) power is sold to (bought from) the PCC.
\end{itemize}
This information is then sent to the DERs and loads, respectively. After that, the PCC solves the optimization problem presented here, obtaining the prices $p_{i,j}$ and discounts $s_{j,i}$ of~\fig{figure:cap_5_market_scenario}.
As in the previous case, different time scales can be used for the electrical and market optimization processes.}

%% file: figures/grid-opt.eps_tex
%% Creator: Inkscape inkscape 0.48.4, www.inkscape.org
%% PDF/EPS/PS + LaTeX output extension by Johan Engelen, 2010
%% Accompanies image file 'grid-opt.eps' (pdf, eps, ps)
%%
%% To include the image in your LaTeX document, write
%%   \input{<filename>.pdf_tex}
%%  instead of
%%   \includegraphics{<filename>.pdf}
%% To scale the image, write
%%   \def\svgwidth{<desired width>}
%%   \input{<filename>.pdf_tex}
%%  instead of
%%   \includegraphics[width=<desired width>]{<filename>.pdf}
%%
%% Images with a different path to the parent latex file can
%% be accessed with the `import' package (which may need to be
%% installed) using
%%   \usepackage{import}
%% in the preamble, and then including the image with
%%   \import{<path to file>}{<filename>.pdf_tex}
%% Alternatively, one can specify
%%   \graphicspath{{<path to file>/}}
%% 
%% For more information, please see info/svg-inkscape on CTAN:
%%   http://tug.ctan.org/tex-archive/info/svg-inkscape
%%
\begingroup%
  \makeatletter%
  \providecommand\color[2][]{%
    \errmessage{(Inkscape) Color is used for the text in Inkscape, but the package 'color.sty' is not loaded}%
    \renewcommand\color[2][]{}%
  }%
  \providecommand\transparent[1]{%
    \errmessage{(Inkscape) Transparency is used (non-zero) for the text in Inkscape, but the package 'transparent.sty' is not loaded}%
    \renewcommand\transparent[1]{}%
  }%
  \providecommand\rotatebox[2]{#2}%
  \ifx\svgwidth\undefined%
    \setlength{\unitlength}{806.108912bp}%
    \ifx\svgscale\undefined%
      \relax%
    \else%
      \setlength{\unitlength}{\unitlength * \real{\svgscale}}%
    \fi%
  \else%
    \setlength{\unitlength}{\svgwidth}%
  \fi%
  \global\let\svgwidth\undefined%
  \global\let\svgscale\undefined%
  \makeatother%
  \begin{picture}(1,0.23490963)%
    \put(0,0){\includegraphics[width=\unitlength]{grid-opt.eps}}%
    \put(0.26527171,0.07391622){\color[rgb]{0,0,0}\makebox(0,0)[lb]{\smash{$G_2$}}}%

    \put(0.04382054,0.07390071){\color[rgb]{0,0,0}\makebox(0,0)[lb]{\smash{$L_2$}}}%

    \put(0.702,0.07390071){\color[rgb]{0,0,0}\makebox(0,0)[lb]{\smash{$L_1$}}}%

    \put(0.885,0.07391622){\color[rgb]{0,0,0}\makebox(0,0)[lb]{\smash{$G_1$}}}%

    \put(0.455,0.20544711){\color[rgb]{0,0,0}\makebox(0,0)[lb]{\smash{PCC}}}%

    \put(0.072,0.20390796){\color[rgb]{0,0,0}\makebox(0,0)[lb]{\smash{$B_4$}}}%

    \put(0.335,0.20406254){\color[rgb]{0,0,0}\makebox(0,0)[lb]{\smash{$B_3$}}}%

    \put(0.605,0.20390796){\color[rgb]{0,0,0}\makebox(0,0)[lb]{\smash{$B_1$}}}%

    \put(0.87,0.20390796){\color[rgb]{0,0,0}\makebox(0,0)[lb]{\smash{$B_2$}}}%

  \end{picture}%
\endgroup%

%% file: notation.tex
%!TEX root = main.tex
In this section, we present the first core contribution of this paper, which consists in an original  optimization modeling of the multi-agent system. 
Our goal is to propose an optimized market model aiming at increasing the DERs monetary revenue and reducing the loads expenses while enforcing the grid electrical efficiency. Given that these are differing optimization objectives, a \mbox{multi-objective} optimization problem is proposed. First, we introduce the mathematical notation that is used throughout the paper. Next, this problem is posed, characterized and transformed into an equivalent convex formulation, which allows for a convenient computation of the optimal Pareto frontier.

\subsection{Notation}
%In this section, the quantities involved in the optimization problem definition and solution and the corresponding notation are introduced.
We now introduce the mathematical notation that is used for the market model throughout the paper, and the multi-objective optimization problem. %that allows to optimally choose the energy pricing and trading policies.

Let $\DGSet$ be the set of active DERs in the grid, $|\DGSet|=G,\ G\in\mathbb{N}$, where $G$ is the cardinality of $\DGSet$. Let $\LSet$ be the set of active loads in the grid, $|\LSet|=L,\ L\in\mathbb{N}$, where $L$ is the cardinality of $\LSet$.

\myparstart{Domains:}
Let
\begin{equation}\label{eq:p_domain}
\PSet = \{\Pmat\in\mathbb{R}_{++}^{G\times L}:\ \Pel{i}{j}\leq P_i,\ \forall \, i\in\DGSet  \}
\end{equation}
be the set of matrices $\Pmat$ whose elements are the unitary prices $\Pel{i}{j}$ that the DERs propose to the loads. 
%The elements of $\PSet$ will be denoted by $\Pmat$. 
The $i,j$ element, for $i = 1,\ldots,G$ and $j = 1,\ldots,L$, of matrix $\Pmat$ is denoted by $\Pel{i}{j}$ and represents the unitary price that $G_i$ proposes to $L_j,\ \forall \, i\in\DGSet,\ \forall \, j\in\LSet$. Let $\Prow{i}$ and $\Pcol{j}$ denote the $i$-th row and the $j$-th column of $\Pmat$, respectively. Moreover, let $P_i, \forall \, i\in\DGSet$ be the PCC imposed maximum unitary price that $G_i$ can propose to the loads.

Let
\begin{equation}\label{eq:h_domain}
\HSet = \{\Hmat\in\mathbb{R}_{+}^{G\times(L+1)}:\ \displaystyle\sum_{j=0}^{L}\Hel{i}{j}= E_i,\ \forall \, i\in\DGSet\}
\end{equation}
be the set of matrices $\Hmat$ representing the amount of power that the DERs can sell to each buyer (the loads or the PCC). 
%The elements of $\HSet$ are denoted by $\Hmat$. 
The $i,j$ element, for $i = 1,\ldots,G$ and $j = 1,\ldots,L$, of matrix $\Hmat$ is denoted by $\Hel{i}{j}$ and represents the amount of power that $G_i$ sells to the buyer $j$ (where $j=0$ denotes the PCC and $j=1,\dots,L$ denotes load $L_j$). With $\Hrow{i}$ and $\Hcol{j}$, we denote the $i$-th row and the $j$-th column of $\Hmat$, respectively.

Let
\begin{equation}\label{eq:d_domain}
\DSet = \{\Dmat\in\mathbb{R}_{+}^{L\times(G+1)}:\ \displaystyle\sum_{j=0}^{G}\Del{i}{j}= D_i,\ \forall \, i\in\LSet\}
\end{equation}
be the set of matrices $\Dmat$ representing the amount of power that the loads can buy from each seller (the DERs or the PCC). 
%The elements of $\DSet$ are denoted by $\Dmat$. 
The $i,j$ element, for $i = 1,\ldots,L$ and $j = 1,\ldots,G$, of matrix $\Dmat$ is denoted by $\Del{i}{j}$ and represents the amount of power that load $L_i$ buys from the seller $j$ (where $j=0$ denotes the PCC and $j=1,\dots,G$ denotes $G_j$). With $\Drow{i}$ and $\Dcol{j}$, we denote the $i$-th row and the $j$-th column of $\Dmat$, respectively.

We aim at bounding the maximum expense from the PCC and, to this end, we introduce a further parameter $\alpha$. Let
\begin{equation}\label{eq:psi_domain}
\SSet = \{\Smat\in\mathbb{R}_{+}^{L\times G}:\ \Sel{i}{j}\leq \alpha \Pel{j}{i} , \forall \, i\in\LSet,\forall \, j\in\DGSet\}
\end{equation}
be the set of matrices $\Smat$ representing the discounts that the PCC applies to the unitary prices that the DERs propose to the loads. We recall that {\textit{the discount policy is meant to drive the electrical grid state toward the optimal one, which is determined by a selected electrical optimization technique}}. 
%The elements of $\SSet$ will be denoted by $\Smat$. 
The $i,j$ element, for $i = 1,\ldots,L$ and $j = 1,\ldots,G$, of matrix $\Smat$ is denoted by $\Sel{i}{j}$ and represents the discount that the PCC is willing to apply to the unitary price $\Pel{i}{j}$ that $G_j$ proposes to $L_i$. Moreover, let $0\leq\alpha\leq 1$ be the PCC defined maximum discount factor, i.e., the PCC is willing to discount at most $(100\alpha)\%$ for each proposed unitary price.

Let $\Hmat \in\HSet$, we define the index sets: $\HrowT{i} = \{k\in\{1,\dots,L\},\ k: \Hel{i}{k} \neq 0\}$ and $\HcolT{j} = \{k\in\{1,\dots,G\},\ k: \Hel{k}{j} \neq 0\}$. These two sets determine the row and column indices, respectively, of the non zero elements of $\Hmat$. 

Similarly, for the demand we define
\begin{equation}
\begin{split}
&\DrowT{i} = \{k\in\{0,\dots,G\}:\ \Del{i}{k}\neq0\}\\
&\DcolT{j} = \{k\in\{1,\dots,L\}:\ \Del{k}{j}\neq0\}.\\
\end{split}
\end{equation}
As above, these two sets respectively determine the row and column indices of the non zero elements of $\Dmat$.

%% file: moo_problem.tex
%!TEX root = main.tex
\subsection{Objective Functions}

Each DER will support the proposed market model, as described in \secref{ssec:cap_5_market_scenario}, only if it guarantees a higher monetary revenue with respect to solely selling its energy to the PCC.
We represent the monetary revenue of the DER $i$ when selling its $E_i$ amount of energy to the loads, as specified by the vector $\Hrow{i}$ and using the unitary prices defined by $\Prow{i}$, by the following equation
\begin{equation}
\label{eq:DER_obj}
\begin{split}
U_i^{\DGSet}(\Prow{i}, \Hrow{i}) = \displaystyle\sum_{j=1}^{L}{\Pel{i}{j}\Hel{i}{j}}+(E_i-\sum_{j=1}^{L}{\Hel{i}{j}}))\gamma_{i}^{\text{PCC}}\\
(\forall \, i\in\DGSet) \, .
\end{split}
\end{equation}
Note that with the second addend we model the fact that all the excess energy from DER $i$ that is not sold to the loads, will be bought by the PCC. This is, in fact, what occurs in current markets and what we also consider here.

In contrast to the DERs behavior, each load will endorse the proposed market model only if it guarantees lower expenses with respect to solely buying energy from the PCC.
Using the demand vector $\Drow{i}$ and the discounted unitary prices \mbox{$\Pcol{i}-\Srow{i}$}, we represent the expense incurred by $L_i$ when buying energy from the DERs as

\begin{equation}
\label{eq:Load_obj}
\begin{split}
U_i^{\LSet}(\Pcol{i}, \Drow{i}, \Srow{i}) & = \displaystyle\sum_{j=1}^{G}{(\Pel{j}{i}-\Sel{i}{j})\Del{i}{j}}\\
& +(D_i-\sum_{j=1}^{G}{\Del{i}{j}})\pi_{i}^{\text{PCC}} \, , \, (\forall \, i\in\LSet) \, .
\end{split}
\end{equation}

We assume that an electrical grid optimization strategy is available and, through it, we obtain the optimal demands $\DrowD{i},\ \forall \, i \in \LSet$ for the loads from an electrical standpoint. These are optimal in the sense that they will drive the grid toward a certain desirable electrical state. Any optimization scheme from the state of the art can be used to obtain these optimal demands, a possible technique is discussed shortly in \secref{sec:sim_setup}. 

Within our market scenario, the PCC \revision{drives the power grid as close as possible to the optimal working
point and, to this end, it enforces a discount} $\Sel{i}{j}$ to each unitary price $\Pel{j}{i}$ that $G_j$ proposes to $L_i$. The effect of the discounts is determined, for each load $L_i$, by computing the squared distance between the chosen demand vector $\Drow{i}$ and the most electrically efficient one $\DrowD{i}$. Such distance is computed through the following equation:
\begin{equation}\label{eq:PCC_obj}
U_i^{\text{PCC}}(\Drow{i}) = ||\Drow{i}-\DrowD{i}||_2^2\quad (\forall \, i\in\LSet) \, .
\end{equation} 
The goal of the PCC is then to determine a discount matrix $\Smat$ that minimizes the distance in Eq.~\ref{eq:PCC_obj}. We recall that each individual discount $\Sel{i}{j}$ is upper bounded by $\Sel{i}{j}\leq\alpha\Pel{j}{i}$.

%a common quantity $0 \leq \alpha \leq 1$ representing the maximum fraction of $\Pel{j}{i}$ that the PCC is willing to discount (i.e., $\Sel{i}{j}\leq\alpha\Pel{j}{i}$).

\subsection{Constraints}

The electrical state of the system induces a set of constraints to account for the physical consistency of the grid. Moreover, an additional set of constraints limits the maximum prices that the DERs can propose to the loads, and the maximum discounts that can be applied to these prices. Next, these constraints are presented and discussed.

We impose that each DER $G_i\in\DGSet$ sells no more than its surplus energy $E_i$, by the following equation
\begin{equation}
\label{eq:Offer_con}
	\displaystyle\sum_{j=1}^{L}\Hel{i}{j} \leq E_i\quad\forall \, i\in\DGSet \, .
\end{equation}
With this constraint we make sure that $G_i$ can not sell more energy than the amount remaining after fulfilling its own needs.

We model the fact that the loads are not equipped with energy storage devices, and hence each load must buy the exact amount of energy needed to fulfill its current power demand. This is implied by the following equation
\begin{equation}
\label{eq:Demand_con}
	\displaystyle\sum_{j=1}^{G}\Del{i}{j} \leq D_i\quad\forall \, i\in\LSet\,,
\end{equation}
and by the second addend of \eq{eq:Load_obj}. \revision{The reason for the inequality in \eq{eq:Demand_con} is because the loads are not required to buy all the energy they need from the DERs, but they can also buy part of it from the PCC, which is referred to as $d_{i,0},\ \forall i\in\LSet$. It follows that $\sum_{j=0}^{G}d_{i,j}=D_i$.}

Furthermore, the amount of energy that DER $G_i\in\DGSet$ is selling to load $L_j\in\LSet$ must be equal to the amount of energy that $L_j$ is buying from $G_i$, i.e., 
\begin{equation}
\label{eq:Consistency_con}
	\Hel{i}{j} = \Del{j}{i} \quad \forall \, i\in\DGSet,\ \forall \, j\in\LSet\,.
\end{equation}
%Imposing these constraints assures that the electrical and economic working points are consistent.

The limits imposed by PCC to the prices that the DERs propose to the loads are modeled by the following constraints
\begin{equation}
\label{eq:Price_con}
	\Pel{i}{j}\leq P_i\quad \forall \, i\in\DGSet,\ \forall \, j\in\LSet\,,
\end{equation}
where the maximum prices act as market regulators, preventing the prices from growing unboundedly and also determine the maximum unitary discount that the PCC is willing to apply.

The maximum fraction of the unitary prices proposed by the DERs that can be discounted by the PCC is set by the constraint
\begin{equation}
\label{eq:Discount_con}
	\Sel{i}{j}\leq\alpha\Pel{j}{i}\quad \forall \, i\in\LSet,\ \forall \, j\in\DGSet\,.
\end{equation}

\subsection{Optimization Problem}
\label{ssec:optimization_problem}

With the objective functions defined in Eqs.~\eqref{eq:DER_obj}-\eqref{eq:PCC_obj} and the constraints of Eqs.~\eqref{eq:Offer_con}-\eqref{eq:Discount_con}, the following multi-objective optimization problem can be formulated:

\begin{subequations}
\label{eq:MOO_problem}
\begin{align}
&\displaystyle\underset{\Pmat,\Hmat,\Dmat,\Smat}{\text{min}} \left[
\begin{array}{lr}
%{1}\over{\displaystyle\UiG} & \forall \, i \in\DGSet\\%
\displaystyle\UiG^{-1} & \forall \, i \in\DGSet\\%
\displaystyle\UiL & \forall \, i \in\LSet\\%
\displaystyle\UiPCC & \forall \, i \in\LSet\\%
\end{array} \right] \label{eq:MOO_problem_a}\\%
&\text{ s.t.} 
\begin{array}{lr} 
\displaystyle\sum_{j=1}^{L}\Hel{i}{j} \leq E_i & \forall \, i \in\DGSet,\forall \, j \in\LSet \\
\displaystyle\sum_{j=1}^{G}\Del{i}{j} \leq D_i & \forall \, i \in\LSet \\
\Pel{i}{j}\leq P_i  & \forall \, i \in\DGSet,\ \forall \, j \in\LSet \\
\Sel{j}{i}\leq\alpha\Pel{i}{j} & \forall \, i \in\DGSet,\ \forall \, j \in\LSet \\
\Hel{i}{j} = \Del{j}{i} & \forall \, i \in\DGSet,\ \forall \, j \in\LSet \, .
\end{array} \label{eq:MOO_problem_b}
\end{align}
\end{subequations}

Solving the multi-objective optimization problem (\ref{eq:MOO_problem}) does not lead to a unique solution, because the objective functions in Eqs.~\eqref{eq:DER_obj}-\eqref{eq:PCC_obj} are contrasting. Hence, simultaneously minimizing these objective functions leads to a set of solutions called \textbf{Pareto Frontier (PF)}. In this paper, we adopt the Pareto multi-objective optimality definition~\cite{VectorOpt}. Next, we investigate the \revision{Pareto-}optimal \revision{(P-optimal)} solution, with particular emphasis on its domain and on the non-convexity of its objective functions. \revision{With $\DmatO$, $\HmatO$, $\SelO{i}{j}$ and $\PelO{i}{j}$ we mean any solution of (\ref{eq:MOO_problem}) resting on the PF.}

\begin{proposition}
\label{prop:optimalDemand}
Consider the optimization problem \eq{eq:MOO_problem} and let $\Pel{j}{i}(1-\alpha)>\pi_{i}^{\text{PCC}}$ for some $j\in\{1,\dots,G\}$ and let $\DrowO{i}$ be the $i$-th row of the \revision{P-}optimal demand matrix $\DmatO\in\DSet$. Then $\DelO{i}{j}=0\, , \, \forall \, \Sel{i}{j} \in \,\, ]0,\alpha\Pel{j}{i}]$ and the \revision{P-}optimal discount value $\SelO{i}{j}$ admits infinite solutions.	
\end{proposition}

\begin{IEEEproof}[Proof]
Let $\Pel{j}{i}(1-\alpha)>\pi_{i}^{\text{PCC}}$ and let $\DrowO{i}$ be the \revision{P-}optimal demand vector for load $i$. If $\DelO{i}{j}\neq 0$, then a new vector $\bm{\bar{d}}_{i,\cdot}$ such that $\bar{d}_{i,j}=0$ and $\bar{d}_{i,0}=\DelO{i}{0}+\DelO{i}{j}$ can be defined. It holds by construction that ${U}_{i}^{\LSet}(\Pcol{i}, \bm{\bar{d}}_{i,\cdot}, \Srow{i})<{U}_{i}^{\LSet}(\Pcol{i}, \DrowO{i}, \Srow{i})$, but this is not possible because $\DrowO{i}$ is \revision{P-}optimal and hence $\DelO{i}{j}=0$.
\end{IEEEproof}

According to \prop{prop:optimalDemand}, if the price that $G_j$ proposes to $L_i$ is still higher than the PCC imposed price, when applying the maximum discount factor $\alpha$, then no feasible discount can make $L_j$ buy power from $G_i$.

\begin{proposition}
\label{prop:optimalAllocation}
Consider optimization problem~\eqref{eq:MOO_problem} and let $\Pel{i}{j}<\gamma^{\text{PCC}}_{i}$ for some $j\in\{1,\dots,L\}$, and let $\HrowO{i}$ be the $i$-th row of the \revision{P-}optimal allocation matrix $\HmatO\in\HSet$. Then $\HelO{i}{j}=0$ and the \revision{P-}optimal discount value $\SelO{j}{i}$ admits infinite solutions.	
\end{proposition}

\begin{IEEEproof}[Proof]
Let $\Pel{i}{j}<\gamma^{\text{PCC}}_{i}$ and let $\HrowO{i}$ be the \revision{P-}optimal allocation vector for DG $i$. If $\HelO{i}{j}\neq 0$, then a new allocation vector $\bm{\bar{h}}_{i,\cdot}$ such that $\bar{h}_{i,j}=0$ and $\bar{h}_{i,0}=\HelO{i}{0}+\HelO{i}{j}$ can be defined. It is true, by construction, that ${U}_{i}^{\DGSet}(\Prow{i}, \bm{\bar{h}}_{i,\cdot})>{U}_{i}^{\DGSet}(\Prow{i}, \HrowO{i})$, but this is not possible because $\HrowO{i}$ is \revision{P-}optimal and hence $\HelO{i}{j}=0$.
\end{IEEEproof}

\prop{prop:optimalAllocation} states that if the revenue that $G_i$ obtains by selling its power to the PCC is greater than the maximum revenue that can be obtained by selling it to $L_j$, then again there is no way for the PCC to enforce the electrical grid efficiency.
To summarize, Propositions~\ref{prop:optimalDemand} and~\ref{prop:optimalAllocation} imply that, in order for the PCC to enforce the grid electrical efficiency through a discount policy, the following conditions must hold $\forall \, i\in\DGSet,j\in\LSet$:

\begin{equation}\label{eq:pricing_conditions}
	\Pel{i}{j}\geq\gamma^{\text{PCC}}_{i}\text{ and }\Pel{i}{j}(1-\alpha)\leq\pi_{j}^{\text{PCC}}
\end{equation}
\revision{We recall that, as discussed in \secref{ssec:cap_5_market_scenario}, the PCC is not allowed to act on $\gamma^{\text{PCC}}_{i}$ and $\pi^{\text{PCC}}_{i}$ as they depend on preexisting contract terms and conditions. Hence, the only way it can act to promote the electrical efficiency of the grid is to allow the DERs to have a higher revenue through selling energy to a specific load, dictated by the joint optimization that is proposed here, as opposed to selling it to any other loads or to the PCC itself. To do this, the PCC discounts the prices that DERs propose. However, to limit its expenses the PCC imposes a maximum discount factor (i.e., $\alpha$) that dictates the limit to within the DERs proposed prices can grow before the loads (being rational agents) stop buying energy from them, i.e., when the price paid to buy energy from a DER is higher than buying it from the PCC ($\Pel{i}{j}(1-\alpha)>\pi_{j}^{\text{PCC}}$).}
\begin{proposition}
\label{prop:optimalJointAllocDem}
Consider optimization problem \eqref{eq:MOO_problem} and let $\HelO{i}{j},\ \DelO{j}{i}$ respectively be the \revision{P-}optimal $i,j$ allocation and $j,i$ demand values (according to the respective indexing). Then, either one of the following holds:
\begin{enumerate}
	\item $\HelO{i}{j} = \DelO{j}{i}=0$ if $\Pel{i}{j}(1-\alpha)>\pi_{j}^{\text{PCC}}$ or $\Pel{i}{j}<\gamma_{i}^{\text{PCC}}$\,;
	\item $\HelO{i}{j} = \DelO{j}{i}\neq 0$ otherwise\,.
\end{enumerate}
\end{proposition}
\begin{IEEEproof}
By considering Propositions~\ref{prop:optimalDemand} and~\ref{prop:optimalAllocation}, and recalling that both the DERs and the loads are rational agents, we see that the only case where it is economically convenient for $G_i$ to sell power to $L_j$ is when it can get a higher revenue than the one it would obtain selling the same amount of power to the PCC. This concludes the proof.
\end{IEEEproof}
\prop{prop:optimalJointAllocDem} follows from Propositions~\ref{prop:optimalDemand} and~\ref{prop:optimalAllocation}. It states that the PCC can enforce the grid efficiency only if the conditions of \eq{eq:pricing_conditions} are met. Moreover, if these conditions are met, the rational behavior for DERs and loads will be to adhere to the discount policy proposed by the PCC and trading energy with the agents guaranteeing higher revenues and smaller expenses for the DERs and loads, respectively.

\begin{proposition}
\label{prop:optimalPrices}
Consider the optimization problem \eq{eq:MOO_problem} and let $\PelO{i}{j}$ be the \revision{P-}optimal $i,j$ price value for the optimization problem. Let $\HelO{i}{j} = \DelO{j}{i}\neq 0$. Then, it must hold that
\begin{enumerate}
	\item $\pi_{j}^{\text{PCC}}<\PelO{i}{j}-\Sel{j}{i}$ for at least one value of $\Sel{j}{i}$\,;
	\item $\PelO{i}{j}>\gamma_{i}^{\text{PCC}}$\,.
\end{enumerate}
\end{proposition}
\begin{IEEEproof}
By considering Propositions~\ref{prop:optimalDemand}-\ref{prop:optimalJointAllocDem} and recalling that loads are rational agents, we see that the only case where $L_j$ will buy power from $G_i$ is when the discounted price proposed by $G_i$ is lower than the price it would pay to the PCC. Moreover, by recalling that DERs are also rational agents, we see that the only case in which $G_i$ will sell power to $L_j$ is the one where its revenue is higher than the one it can get from the PCC. This concludes the proof.
\end{IEEEproof}
\prop{prop:optimalPrices} descends from \prop{prop:optimalJointAllocDem}. It states that, in order for the DERs and loads to adhere to the proposed model, the discounts (limited to a fraction $\alpha$ of the proposed prices) must meet the rational behavior of the trading agents. 

\revision{Propositions~\ref{prop:optimalDemand}-\ref{prop:optimalPrices} characterize the PF of problem~\eqref{eq:MOO_problem}. From these propositions, it follows that, by construction, no solution on the PF can lead to situations where 1) some DERs experience smaller revenues with respect to the case where all the energy is sold to the PCC or 2) some loads experience higher expenses with respect to the case where all the energy is bought from the PCC. Hence, network agents deciding to adopt the proposed optimized market model have no reason not to accept the trading strategy resulting from the solution of the optimization problem~\eqref{eq:MOO_problem}.}

In the following proposition, we show that the domains defined through Eqs.~(\ref{eq:p_domain})-(\ref{eq:psi_domain}) are convex.

\begin{proposition}
The sets defined in Eqs.~(\ref{eq:p_domain})-(\ref{eq:psi_domain}) are convex with respect to the matrix sum operation.
\end{proposition}

\begin{IEEEproof}[Proof]
\label{proof:convPi}
Let $\bm{P_1},\bm{P_2}\in\PSet$ and $0\leq\theta\leq1$. Let $\bm{P_3} = \theta \bm{P_1}+(1-\theta)\bm{P_2}$, then
\begin{equation}
\displaystyle\sum_{j=1}^{L}{{p_3}}_{i,j} = \theta \displaystyle\sum_{j=1}^{L}{p_1}_{i,j} + (1-\theta) \displaystyle\sum_{j=1}^{L}{p_2}_{i,j} \, .
\end{equation}
Since $\bm{P_1},\bm{P_2}\in\PSet$, it holds true that
\begin{equation}
\forall \, i\in\DGSet,\ \displaystyle\sum_{j=1}^{L}{p_1}_{i,j}\leq P_i \,\, \textrm{and} \,\,\displaystyle\sum_{j=1}^{L}{p_2}_{i,j}\leq P_i
\end{equation}
hence
\begin{equation}
\displaystyle\sum_{j=1}^{L}{p_3}_{i,j} \leq \theta P_i + (1-\theta) P_i = P_i,\ \forall \, i\in\DGSet  \, ,
\end{equation}
thus $\bm{P_3}\in\PSet$.
\end{IEEEproof}
The convexity of sets $\HSet$, $\DSet$ and $\SSet$ can be shown by similar arguments.

Next, we show that although the domains are convex the multi-objective optimization problem of \eq{eq:MOO_problem} is not.

\begin{proposition}\label{prop:Non_Convexity}
Optimization problem \eqref{eq:MOO_problem} is not convex. 
\end{proposition}
%\begin{IEEEproof}[Proof of proposition \ref{prop:Non_Convexity}]
\begin{IEEEproof}[Proof]
In order to prove the non convexity of \eq{eq:MOO_problem} it is sufficient to show that one of its objective functions is not convex. Considering $\UiG$, since  $\UiG$ is twice differentiable in its domain the Hessian matrix $\bm{\Phi}_{\UiG}$ can be computed:
\begin{equation}
\bm{\Phi}_{\UiG} = \left[
\begin{array}{c|c}
\bm{A} & \bm{B} \\
\hline
\bm{B} & \bm{A}
\end{array}
\right]\,,
\end{equation}
where $\bm{A} \in\{0\}^{L\times L}$ and $\bm{B}$ is the $L \times L$ identity matrix. $\bm{\Phi}_{\UiG}$ is a permutation matrix. Let $\bm{z}\in\mathbb{R}^{2L}$ and let $\bm{z}_1,\bm{z}_2 \in \mathbb{R}^{L}\ :\ \bm{z}^T = [\bm{z}_1^T \bm{z}_2^T]$, then
\begin{equation}
	\bm{z}^T \bm{\Phi}_{\UiG}\bm{z} = [\bm{z}_2^T\bm{z}_1^T]\bm{z}\,,
\end{equation}
and hence $\bm{\Phi}_{\UiG}$ is not positive semidefinite nor it is negative semidefinite.
\end{IEEEproof}
According to \prop{prop:Non_Convexity}, solving problem \eqref{eq:MOO_problem} with standard convex multi-objective solution methods could not lead to the actual PF. In the following subsection, a transformation of problem \eqref{eq:MOO_problem} is proposed, which establishes an equivalent convex optimization problem whose solutions are the same as those of problem \eqref{eq:MOO_problem}. The convexity of the new problem allows the application of standard solution methods. 

\subsection{Geometric Programming Formulation}

Since the DERs' and loads' objective functions can be expressed in posynomial form, part of the non-convex multi-objective minimization problem \eqref{eq:MOO_problem} can be formulated as a geometric programming problem~\cite{CV-BOYD,GP-BOYD,NL-MOBJ}. Next, the steps leading to this transformation will be presented and discussed.

First, we can express \eq{eq:DER_obj} in the form
\begin{equation}
\label{eq:almost_posy_DER_obj}
\UiG = \displaystyle\sum_{j=1}^{L}{\Pel{i}{j}\Hel{i}{j}}+\Hel{i}{0}\gamma_{i}^{\text{PCC}}\quad\forall \, i\in\DGSet\,.
\end{equation}
Now, we consider the following definitions: 
\begin{equation}
\alpha_{j}^{\DGSet} =%
\begin{cases}
0&\text{if } j=0\\
1&\text{otherwise}
\end{cases}
\, , \,\,  c_{ij}^{\DGSet} = %
\begin{cases}
\gamma_i^{\text{PCC}}&\text{if } j=0\\
1&\text{otherwise}
\end{cases}
\end{equation}
and let $\Pel{i}{0}\in\mathbb{R}\ \forall \, i \in\DGSet$. Then, \eq{eq:almost_posy_DER_obj} can be rewritten as a posynomial function:
%Now, extending the definition of $\Pmat$ as 
%\begin{equation}
%\Pel{i}{0}  = \gamma_i^{\text{PCC}} \forall \, i \in\DGSet
%\end{equation}
%then, \eq{eq:almost_posy_DER_obj} can be rewritten as a posynomial function:
\begin{equation}
\label{eq:posy_DER_obj}
	\UiG = \displaystyle\sum_{j=0}^{L}{c_{ij}^{\DGSet}\Pel{i}{j}^{\alpha_{j}^{\DGSet}}\Hel{i}{j}} \, , \, \forall \, i \in \DGSet\,.
\end{equation}

Similarly, \eq{eq:Load_obj} can be re-formulated as
\begin{equation}
\label{eq:almost_posy_Load_obj}
	\UiL = \displaystyle\sum_{j=1}^{G}{(\Pel{j}{i}-\Sel{i}{j})\Del{i}{j}}+\Del{i}{0}\pi_i^{\text{PCC}} \, , 
\end{equation}
$\forall \, i\in\LSet$. 
If we define
\begin{equation}
\alpha_{j}^{\LSet} =%
\begin{cases}
0&\text{if } j=0\\
1&\text{otherwise}
\end{cases}
\textrm{ and } c_{ij}^{\LSet} = %
\begin{cases}
\pi_i^{\text{PCC}}&\text{if } j=0\\
1&\text{otherwise} \, ,
\end{cases}
\end{equation}
\eq{eq:almost_posy_Load_obj} can also be formulated as a posynomial function:
%If we define 
%\begin{equation}
%\PelPrime{0}{i} =  \begin{cases}
%\pi_i^{\text{PCC}} & j= 0 \\
%\Pel{j}{i}-\Sel{i}{j} &  {\rm otherwise} 
%\end{cases}
%\end{equation}
%$\forall \, i\in\LSet$, then \eq{eq:almost_posy_Load_obj} can also be formulated as a posynomial function:
\begin{equation}\label{eq:posy_LOAD_obj}
	\UiLprime = \displaystyle\sum_{j=0}^{G}{c_{ij}^{\LSet}\PelPrime{j}{i}^{\alpha_{j}^{\LSet}}\Del{i}{j}} \, , \, \forall \, i \in \LSet\,.
\end{equation}
%where $\PelPrime{j}{i} = \Pel{j}{i}-\Sel{i}{j}$.

If we apply the geometric programming transformation detailed in~\cite{GP-BOYD}, \eq{eq:posy_DER_obj}  and \eq{eq:posy_LOAD_obj} can be transformed into convex functions:
\begin{equation}\label{eq:cvx_DER_obj}
\cvxUiG = \displaystyle\sum_{j\in\HrowT{i}}{e^{\alpha_{j}^{\DGSet}\log{\Pel{i}{j}}+\log{\Hel{i}{j}}+\log{c_{ij}^{\DGSet}}}}\,,
\end{equation}
\begin{equation}\label{eq:cvx_Load_obj}
\cvxUiL = \displaystyle\sum_{j\in\DrowT{i}}{e^{\alpha_{j}^{\LSet}\log{\PelPrime{j}{i}}+\log{\Del{i}{j}}+\log{c_{ij}^{\LSet}}}}\,.
\end{equation}
Based on Eqs.~(\ref{eq:cvx_DER_obj}) and (\ref{eq:cvx_Load_obj}), the non-convex multi-objective optimization problem \eqref{eq:MOO_problem} can be transformed into a convex multi-objective optimization problem:
\begin{subequations}
\label{eq:CVXProblem}
\begin{align}
&\displaystyle\min_{\Pmat,\Hmat,\Dmat,\Smat} \left[
\begin{array}{lr}
-\log{\left(\displaystyle\cvxUiG\right)} & \forall \, i\in\DGSet\\%
\displaystyle\log{\left(\cvxUiL\right)} & \forall \, i\in\LSet\\%
\displaystyle\UiPCC & \forall \, i\in\LSet\\%
\end{array} \right] \\%
& \text{ s.t.}
\begin{array}{lr} 
%\displaystyle\sum_{j=1}^{L}\Hel{i}{j} \leq E_i & \forall \, i\in\DGSet,\forall \, j\in\LSet \\
%\displaystyle\sum_{j=1}^{G}\Del{i}{j} \leq D_i & \forall \, i\in\LSet \\
%\Pel{i}{j}\leq P_i  & \forall \, i\in\DGSet,\ \forall \, j\in\LSet \\
%\Sel{j}{i}\leq\alpha\Pel{i}{j} & \forall \, i\in\DGSet,\ \forall \, j\in\LSet \\
%\Hel{i}{j} = d_{ji} & \forall \, i\in\DGSet,\ \forall \, j\in\LSet \\ 
\text{constraints in \eq{eq:MOO_problem_b}}\\
\PelPrime{j}{i} = \Pel{j}{i}-\Sel{i}{j} & \forall \, i\in\LSet,\ \forall \, j\in\DGSet \, .
\end{array}
\end{align}
\end{subequations}
Note that this optimization problem is equivalent to \eqref{eq:MOO_problem} in the sense that the \revision{P-}optimal solutions of problem \eqref{eq:CVXProblem} are identical to those of problem \eqref{eq:MOO_problem}. %We are now ready to establish a solution method to problem \eqref{eq:MOO_problem} in the next section. 

\subsection{Solution}

Given the convexity of optimization problem \eqref{eq:CVXProblem}, whose solutions are identical to \eqref{eq:MOO_problem}, its solution can be obtained through standard convex solvers. Since problem \eqref{eq:CVXProblem} is convex, the duality gap is zero and the Karush-Kuhn-Tucker (KKT) optimality conditions can be applied to the scalarized form of problem (26). To do so, let $\bm{\lambda} = [\lambda_1, \lambda_2, \dots, \lambda_{G+2L}]^T \in[0,1]^{G+2L}\ :\ \sum_{i=1}^{G+2L}\lambda_i=1$, then the scalarized objective function is
\begin{equation}
\label{eq:scalarized_CVXObj}
\begin{aligned}
	U(\Pmat,\Hmat,\Dmat,\Smat)&= - \sum_{i=1}^{G}\lambda_i \log \left (\cvxUiG \right )\\
	&+\sum_{i=1}^{L}\lambda_{i+G}\log \left (\cvxUiL \right )\\
	&+\sum_{i=1}^{L}\lambda_{i+G+L}\UiPCC \, .
\end{aligned}
\end{equation}
Then, the scalarized convex minimization problem can be defined
\begin{subequations}
\label{eq:scalarized_CVXProblem}
\begin{align}
&\displaystyle\min_{\Pmat,\Hmat,\Dmat,\Smat}
U(\Pmat,\Hmat,\Dmat,\Smat)\\%
&\text{ s.t.}
\begin{array}{lr} 
\displaystyle\sum_{j=1}^{L}\Hel{i}{j} \leq E_i & \forall \, i\in\DGSet,\forall \, j\in\LSet \,\ \\
\displaystyle\sum_{j=1}^{G}\Del{i}{j} \leq D_i & \forall \, i\in\LSet \,\ \\
\Pel{i}{j}\leq P_i  & \forall \, i\in\DGSet,\ \forall \, j\in\LSet \,\ \\
\Sel{j}{i}\leq\alpha\Pel{i}{j} & \forall \, i\in\DGSet,\ \forall \, j\in\LSet \,\ \\
\Hel{i}{j} = d_{ji} & \forall \, i\in\DGSet,\ \forall \, j\in\LSet \,\ \\ 
\PelPrime{j}{i} = \Pel{j}{i}-\Sel{i}{j} & \forall \, i\in\LSet,\ \forall \, j\in\DGSet \, .
\end{array}
\end{align}
\end{subequations}
\eq{eq:scalarized_CVXProblem} is a standard convex minimization problem, hence, if the problem is feasible, a \revision{P-}optimal solution is guaranteed to exist $\forall \, \bm{\lambda}\in[0,1]^{G+2L}\ :\ \sum_{i=1}^{G+2L} \lambda_i =1$. The \textbf{Pareto frontier} is the set of all the \revision{P-}optimal solutions to problem \eqref{eq:scalarized_CVXProblem} obtained for every possible weight vector $\bm{\lambda}$. We recall that all the points in the Pareto frontier are equally \revision{P-}optimal \revision{(in the sense that all these solutions yield the same value of the scalarized objective function)}, it is up to the decision maker \revision{(i.e., the PCC)} to determine the particular weight vector satisfying her/his own needs. In the following results, $\bm{\lambda}$ was heuristically chosen to guarantee that the best electrical working point is reached for low values of the discount factor $\alpha$. The rationale is to drive the grid toward the wanted electrical operating point by maintaining the expenses incurred by the PCC low. \revision{As discussed in \secref{ssec:optimization_problem}, all the solutions of problem~\eqref{eq:scalarized_CVXProblem} lying on the PF guarantee economic benefits to all the DERs and the loads participating in the proposed market. Hence, regardless of the specific $\bm{\lambda}$ that is selected, all the network agents will benefit from embracing the corresponding \revision{P-}optimal solution. The choice of $\bm{\lambda}$ determines the entity of the benefit that each individual agent will achieve thanks to the optimization process. In the proposed model, it is the PCC that selects, during each optimization phase, the most suitable $\bm{\lambda}$ with respect to each specific scenario. For example, in certain scenarios it might be appropriate to specifically enforce the electrical optimization in some regions of the grid. In such a case, the components of $\bm{\lambda}$ associated with agents connected to these particular regions will be higher than the others. By doing so, minimizing these functions will have a greater impact on the overall minimization with respect to the functions of agents in other portions of the grid. We remark that the choice of the vector $\bm{\lambda}$ determines \revision{the particular solution on the PF}, but it is not part of the optimization process.}

\subsection{Distributed Solution}
\label{ssec:distributed}

\revision{Here, we present a decentralized solution of problem \eqref{eq:scalarized_CVXProblem} exploiting the alternating direction method of multipliers (ADMM)~\cite{CV-BOYD}. Consider the function $\cvxUiG$ of \eq{eq:cvx_DER_obj}, and, for the ease of notation, define 
\begin{equation}
y_{i,j} = {e^{\alpha_{j}^{\DGSet}\log{\Pel{i}{j}}+\log{\Hel{i}{j}}+\log{c_{ij}^{\DGSet}}}}\ \forall j\in\HrowT{i}\\.
\end{equation}
Then, \eq{eq:cvx_DER_obj} can be rewritten as 
\begin{equation}
\cvxUiG = \displaystyle\sum_{j\in\HrowT{i}}{y_{i,j}} = {U'}_{i}^{\DGSet}(\bm{y}_{i,\cdot}) \, ,
\end{equation}
and, for the Jensen's inequality:
\begin{equation}
\log{\left({{1}\over{|\HrowT{i}|}}{U'}_{i}^{\DGSet}(\bm{y}_{i,\cdot})\right)}\geq {{1}\over{|\HrowT{i}|}}\displaystyle\sum_{j\in\HrowT{i}}{\log{(y_{i,j})}}\, ,
\label{eq:jens_trans}
\end{equation}
$\forall i\in\DGSet$.}

\begin{proposition}
\label{prop:argmax_conservation}
\revision{
The feasible vector
\begin{equation}
\bm{y}^{\bm{*}}_{i,\cdot} = \underset{\bm{y}_{i,\cdot}}{\textrm{argmax}}\left(\displaystyle\sum_{j\in\HrowT{i}}{\log{(y_{i,j})}}\right)\, ,
\end{equation}
subject to the constraints of problem \eqref{eq:CVXProblem}, is unique and maximizes ${U'}_{i}^{\DGSet}(\bm{y}_{i,\cdot})$.
}
\end{proposition}

\begin{IEEEproof}[Proof]
\revision{
Since the logarithm is a strictly concave function, the sum of logarithms is strictly concave. Hence, $\bm{y}^{\bm{*}}_{i,\cdot}$ is unique. Since $\bm{y}^{\bm{*}}_{i,\cdot}$ is unique, it follows that $\forall \text{ feasible } \bm{y}_{i,\cdot}\neq\bm{y}^{\bm{*}}_{i,\cdot}\ y_{i,j}\leq y^{*}_{i,j} \forall j\in\HrowT{i}$. Then, 
\begin{equation}
\forall \text{ feasible } \bm{y}_{i,\cdot}\neq\bm{y}^{\bm{*}}_{i,\cdot}\ \sum_{j\in\HrowT{i}}{y_{i,j}}\leq \sum_{j\in\HrowT{i}}{y^{*}_{i,j}}\, .
\end{equation} 
This concludes the proof.
}
\end{IEEEproof}
\revision{
As a consequence of Proposition~\ref{prop:argmax_conservation}, the feasible solution that maximizes the right-hand side of \eq{eq:jens_trans}, also maximizes the original objective function $\log{\left({U'}_{i}^{\DGSet}(\bm{y}_{i,\cdot})\right)}$. Define 
\begin{equation}
\dot{U}^{\DGSet}_{i}(\bm{y}_{i,\cdot}) = \displaystyle\sum_{j\in\HrowT{i}}{\log{(y_{i,j})}}\, ,\ \forall i\in\DGSet\, .
\end{equation}
Similarly, define 
\begin{equation}
\dot{U}^{\LSet}_{i}(\bm{z}_{i,\cdot}) = \displaystyle\sum_{j\in\DrowT{i}}{\log{(z_{i,j})}}\, ,\ \forall i\in\LSet\, ,
\end{equation}
where $z_{i,j} = e^{\alpha_{j}^{\LSet}\log{\PelPrime{j}{i}}+\log{\Del{i}{j}}+\log{c_{ij}^{\LSet}}}$. For the same argument of Proposition~\ref{prop:argmax_conservation}, the vector that minimizes $\dot{U}^{\LSet}_{i}(\bm{z}_{i,\cdot})$ is unique and also minimizes $\log{\left({U'}_{i}^{\LSet}(\bm{y}_{i,\cdot})\right)}$. By substituting ${U'}_{i}^{\DGSet}(\bm{y}_{i,\cdot})$ and ${U'}_{i}^{\LSet}(\bm{z}_{i,\cdot})$ with $\dot{U}^{\DGSet}_{i}(\bm{y}_{i,\cdot})$ and $\dot{U}^{\LSet}_{i}(\bm{z}_{i,\cdot})$, respectively, in \eq{eq:scalarized_CVXObj}, the objective function of problem~\eqref{eq:scalarized_CVXProblem} becomes separable with respect to the interactions between any pair of network agents. In the following, we show how to exploit this separability to express problem~\eqref{eq:scalarized_CVXProblem} as a general form consensus with regularization~\cite{BOYD-ADMM}. Consider a smart grid with $K>0$ branches departing from the PCC. Define region $R_k$ as the set of grid agents (i.e., loads and DERs) connected to the PCC through the $k$-th branch. Then, the DER set $\DGSet$ can be partitioned into $K$ subsets $\DGSet^{R_k},\ k=1,\dots,K$ such that $\DGSet^{R_1}\cup\dots\cup\DGSet^{R_K}=\DGSet$, and $\DGSet^{R_1}\cap\dots\cap\DGSet^{R_K}=\emptyset$ where the set $\DGSet^{R_k}$ is the set of DERs belonging to region $k$. Likewise, the set of loads $\LSet$ can be partitioned into $K$ subsets $\LSet^{R_1},\dots,\LSet^{R_K}$. According to these partitions, and assuming that $g\in\DGSet^{R_k}$ and $l\in\LSet^{R_k}$, $\dot{U}^{\DGSet}_{g}(\bm{y}_{g,\cdot})$ and $\dot{U}^{\LSet}_{l}(\bm{z}_{l,\cdot})$ can be rewritten as:
\begin{equation}
\label{eq:separated_DER_Obj}
\dot{U}^{\DGSet}_{g}(\bm{y}_{g,\cdot}) = \displaystyle \!\!\!\!\!\! \sum_{j\in\HrowT{g}\cap\DGSet^{R_k}}{ \!\!\!\!\!\! \log{(y_{g,j})}} + \!\!\!\!\!\!\!\! \displaystyle\sum_{j\in\HrowT{g}\cap(\DGSet\setminus\DGSet^{R_k})}{ \!\!\!\!\!\!\!\!\!\! \log{(y_{g,j})}}\, ,
\end{equation}
and
\begin{equation}
\label{eq:separated_L_Obj}
\dot{U}^{\LSet}_{l}(\bm{z}_{l,\cdot}) = \displaystyle \!\!\!\!\!\!  \sum_{j\in\DrowT{l}\cap\LSet^{R_k}}{ \!\!\!\!\!\! \log{(z_{l,j})}} + \!\!\!\!\!\!\!\! \displaystyle\sum_{j\in\DrowT{l}\cap(\LSet\setminus\LSet^{R_k})}{ \!\!\!\!\!\!\!\!\!\! \log{(z_{l,j})}}\, ,
\end{equation}
respectively. For the ease of notation, \eq{eq:separated_DER_Obj} can be rewritten as 
\begin{equation}
\dot{U}^{\DGSet}_{g}(\bm{y}_{g,\cdot}) = \dot{U}^{\DGSet^{R_k}}_{g}(\bm{y}^{k}_{g,\cdot}) + \dot{U}^{\DGSet\setminus\DGSet^{R_k}}_{g}(\bm{y}^{\breve{k}}_{g,\cdot}) \, ,
\end{equation}
where the first term in the RHS corresponds to the first sum in the RHS of \eq{eq:separated_DER_Obj}, and the second term corresponds to the second sum in the RHS of \eq{eq:separated_DER_Obj}, where $\bm{y}^{k}_{g,\cdot}$ models interactions between $g\in\DGSet^{R_k}$ and agents in the same region $R_k$, while $\bm{y}^{\breve{k}}_{g,\cdot}$ models interactions across regions.
The same decomposition can be applied to loads ($\LSet$) and the PCC, leading to:
\begin{equation}
\begin{aligned}
\dot{U}^{\LSet}_{l}(\bm{z}_{l,\cdot}) = \dot{U}^{\LSet^{k}}_{l}(\bm{z}^{k}_{l,\cdot}) + \dot{U}^{\LSet\setminus\LSet^{k}}_{l}(\bm{z}^{\breve{k}}_{l,\cdot}) \, , \\
U^{\text{PCC}}_l(\bm{d}_{l,\cdot}) = U^{\text{PCC}}_l(\bm{d}^{k}_{l,\cdot}) + U^{\text{PCC}}_l(\bm{d}^{\breve{k}}_{l,\cdot}) \, .
\end{aligned}
\end{equation}
The separability of all the objective functions allows defining a scalarized optimization problem whose solution is equivalent to the one of problem~\eqref{eq:scalarized_CVXProblem}. This can be done by defining the new scalarized objective function as
\begin{equation}
\begin{aligned}
\dot{U}(\Pmat,\Hmat,\Dmat,\Smat) = \displaystyle\sum_{k=1}^{K}\left(\sum_{i\in\DGSet^{R_k}}\lambda_{i}\dot{U}^{\DGSet^{R_k}}_{i}(\bm{y}^{k}_{i,\cdot})\right. &+\\
\left. \sum_{i\in\LSet^{R_k}}\left(\lambda_{i+G}\dot{U}^{\LSet^{k}}_{l}(\bm{z}^{k}_{i,\cdot})+\lambda_{i+2G}U^{\text{PCC}}_l(\bm{d}^{k}_{i,\cdot})\right.\right) &+ \\
\sum_{k=1}^{K}\left(\sum_{i\in\DGSet}\lambda_{i}\dot{U}^{\DGSet\setminus\DGSet^{R_k}}_{i}(\bm{y}^{\breve{k}}_{i,\cdot})\right. &+\\
\left. \sum_{i\in\LSet^{R_k}}\left(\lambda_{i+G}\dot{U}^{\LSet\setminus\LSet^{k}}_{i}(\bm{z}^{\breve{k}}_{i,\cdot})+\lambda_{i+2G}U^{\text{PCC}}_i(\bm{d}^{\breve{k}}_{i,\cdot})\right.\right)\, ,
\end{aligned}
\end{equation}
which can be rewritten in compact form as:
\begin{equation}
\label{eq:final_obj_admm}
\dot{U}(\Pmat,\Hmat,\Dmat,\Smat) = \sum_{i=1}^{K}\dot{U}^{R_k}(\bm{V}^k) + \dot{U}^{\rm cross}(\bm{W})\, ,
\end{equation}
where $\bm{V}^k$ is a vector containing the portions of $\bm{P},\ \bm{H},\ \bm{D},\ \bm{S}$ pertaining only agents in region $R_k$, $\dot{U}^{\rm cross}(\bm{W})$ contains the \mbox{cross-terms} modeling the interactions across regions, while $\bm{W} = (\bm{P},\ \bm{H},\ \bm{D},\ \bm{S})$. For construction, $\bm{V}^k$ is independent of any other $\bm{V}^r:\ r\neq k$. Replacing \eq{eq:scalarized_CVXObj} with \eq{eq:final_obj_admm} into problem~\eqref{eq:scalarized_CVXProblem} and adding the constraints that guarantee that each $\bm{V}^k$ is consistent with $\bm{W}$, generates a new problem that is equivalent to the previous one but is posed as a general form consensus with regularization. This problem can be efficiently solved in a decentralized fashion using ADMM, as shown in~\cite{BOYD-ADMM,SINDRI-FISCHIONE}. We recall that, since the new considered problem is convex, ADMM is guaranteed to converge to the unique \revision{P-}optimal solution (for each given $\bm{\lambda}$). Hence, the trading strategies that are obtained through this decentralized approach are the same as those obtained through the centralized one.
}

In the following section, we present numerical results to characterize the \revision{P-}optimal solutions of our original multi-objective optimization problem \eqref{eq:MOO_problem} based on the established equivalent scalarized problem \eqref{eq:scalarized_CVXProblem}.

%% file: sim_setup.tex
%!TEX root = main.tex
In this section, we present the electrical grid topology and the electrical scenarios, in terms of power demand at the loads and surplus energy at the DERs, that will be considered for our subsequent performance analysis of Section~\ref{sec:results}. 

We consider the electrical grid of \fig{figure:cap_5_electrical_network} as a case study. To determine the optimal power demand matrix, i.e., $\DrowD{i},\ \forall \, i \in \LSet$, we selected the Current Based Surround Control algorithm (CBSC)~\cite{SurroundControl}. The reason why the CBSC algorithm has been chosen is twofold. On the one hand, it drives the grid toward its theoretically optimal working point, and hence it allows the assessment of the optimization process ability to drive the power grid toward its maximum electrical efficiency. On the other hand, the communication infrastructure requirements needed to implement CBSC are the same needed to implement the proposed optimization strategy. Both techniques, indeed, require that each node is equipped with a smart metering device (to determine the exact power availability, power demand and line impedance) and a transceiver (to communicate the measured data and implement control actions). Lastly, CBSC was proven to be very efficient and to lead to optimal results in terms of power loss minimization along the distribution lines. Still, we also recall that other optimization techniques can be used in combination with our optimization framework.

CBSC groups the nodes into {\it clusters}. Clusters are defined by checking, for any pair of DERs, whether their connecting path includes any other DER or the PCC. If this is not the case, a {\it cluster} is defined as the set containing the two DERs, the associated nodes, and all the nodes between them in the electrical network topology. For each cluster, the DER that is closest to the PCC is elected as the cluster head (CH). In the case when one of the two DERs in the cluster is the PCC, this is elected as the CH (i.e., we assume that the PCC has better communication and computational resources with respect to the other nodes). The current injected for optimization purposes is scaled by a real factor $0 \leq \xi \leq 1$. If we refer to $I_C$ as the total current demand in the cluster, the currents injected by the two DESs therein are $\xi I_C$ and $(1-\xi)I_C$. The parameter $\xi$ is determined for each cluster according to the instantaneous power demand from its loads and their branch impedances. Hence, this technique requires that every node is a smart node (i.e., equipped with metering, communication and control capabilities).

According to CBSC, the optimal power allocation matrix, for the considered grid topology, is
\begin{equation}
\label{eq:optimal_demand_matrix}
\bm{D^{\diamond}} = \left[
\begin{array}{c c}
		\displaystyle B_2 {{D_{1}}\over{B_1+B_2}} & 0 \\
		0 & D_{2}
\end{array}	\right] \, , 
\end{equation}
where with $B_i$, $i=1,\dots,4$ we indicate the length of the distribution lines, see table \tab{tab:branch_lengths}, whereas $D_1$ and $D_2$ are the power demands associated with the two loads $L_1$ and $L_2$, which have been set to $D_1=D_2=100$~kW.
According to \eq{eq:optimal_demand_matrix} and \tab{tab:branch_lengths}, the optimal power demand matrix is:
\begin{equation}
	\bm{D^{\diamond}} = \left[
	\begin{array}{c c}
		$50$ \text{kW} & $0$ \\
		$0$ & $100$ \text{kW}
	\end{array} \right]\,.
\end{equation}
Given the optimal power demand matrix of \eq{eq:optimal_demand_matrix}, three electrical scenarios have been considered: 1) the first scenario is referred to as \textit{tight power offer} as it addresses the case where the individual surplus energy for each DER equals the total energy that it should inject according to CBSC. 2) The second scenario is referred to as \textit{unbalanced tight power offer}. Here, the total surplus energy equals the optimal one that should be injected, but the individual surplus energy does not match that dictated by CBSC. In this case, the optimal electrical grid conditions can not be reached. 3) The third is referred to as \textit{loose power offer}. This scenario considers the case where the total surplus energy exceeds the total energy demand.  

\begin{table}
\centering
	\caption{Distribution lines length in meters}\label{tab:branch_lengths}
	\begin{tabular}{|c|c|c|c|}
	\hline
	$B_1$ & $B_2$ & $B_3$ & $B_4$ \\
	\hline
	$50$ m & $50$ m & $45$ m & $90$ m \\
	\hline
	\end{tabular}
\end{table}

\begin{table}
\centering
\caption{DERs surplus energy ($E_i$)}\label{tab:power_availability}
\begin{tabular}{|c|c|c|c|}
\hline
 & \text{Scenario 1} & \text{Scenario 2} & \text{Scenario 3} \\
 \hline
$\textbf{G}_1$ & 50\text{kW} & 60\text{kW} & 100\text{kW} \\
\hline
$\textbf{G}_2$ & 100\text{kW} & 90\text{kW} & 100\text{kW} \\
\hline
\end{tabular}
\end{table}

The DERs surplus energy for each considered scenario is shown in \tab{tab:power_availability}.

%% file: results.tex
%!TEX root = main.tex
In this section, we discuss the \revision{P-}optimal solutions obtained through the proposed optimization approach to the case study of \secref{sec:sim_setup}.

For each scenario, the performance of the optimization process has been assessed using the following metrics:
\begin{itemize}
	\item the DERs gain, obtained as 
		\begin{equation*}
		\textrm{gain} = {{\textrm{revenue}_{\rm opt}-\textrm{revenue}_{\rm no}}\over{\textrm{revenue}_{\rm no}}} \times 100 \, ,
		\end{equation*}
		where ``revenue$_{\rm opt}$'' is the aggregated revenue of DERs when our joint optimization is used, whereas ``revenue$_{\rm no}$'' corresponds to the DERs aggregated revenues in the \mbox{non-optimized} case, i.e., where the surplus power is entirely sold to the PCC;
	\item the loads gain, obtained as 
		\begin{equation*}
		\textrm{gain} = {{\textrm{expense}_{\rm no}-\textrm{expense}_{\rm opt}}\over{\textrm{expense}_{\rm no}}} \times 100 \, ,
		\end{equation*}
		where ``expense$_{\rm opt}$'' is the aggregated expense of loads when our joint optimization is used, whereas ``expense$_{\rm no}$'' corresponds to the loads aggregated expense in the \mbox{non-optimized} case, i.e., where the needed power is entirely bought from the PCC;
	\item the achieved electrical efficiency with respect to the theoretical optimal working point, achieved by CBSC.
\end{itemize}
\revision{Moreover, we set $\gamma_i^{\text{PCC}} = 20,\ \forall i\in\DGSet$ and $\pi_i^{\text{PCC}} = 50,\ \forall i\in\LSet$. This models the real-world scenario where the PCC buys energy for less than what it sells it for. Given these values, the prices $p_{i,j}$ can range from $50$ to $50(1+\alpha)$. To select one particular solution on the PF we chose a specific vector $\bm{\lambda}$. This vector has the following properties:
\begin{itemize}
\item functions of the same class have the same weight, i.e., $\lambda_1=\lambda_2$ (for the DERs), $\lambda_3=\lambda_4$ (for the loads), and $\lambda_5=\lambda_6$ (for the PCC);
\item $\lambda_1 = 0.3\lambda_3$;
\item $\lambda_1 = 0.1\lambda_5$.
\end{itemize}
By doing so, we give the higher weight to the functions enforcing the electrical efficiency, while the revenue of the DERs becomes less important in the global optimization process.
}
Next, we discuss the performance for the three scenarios identified in \secref{sec:sim_setup}. 

\subsection{Scenario 1: Tight Power Offer}

In the tight power offer scenario, $G_1$ and $G_2$ sell the exact amount of power dictated by the PCC.

\begin{figure}[t]
\centering
\includegraphics[scale=0.7]{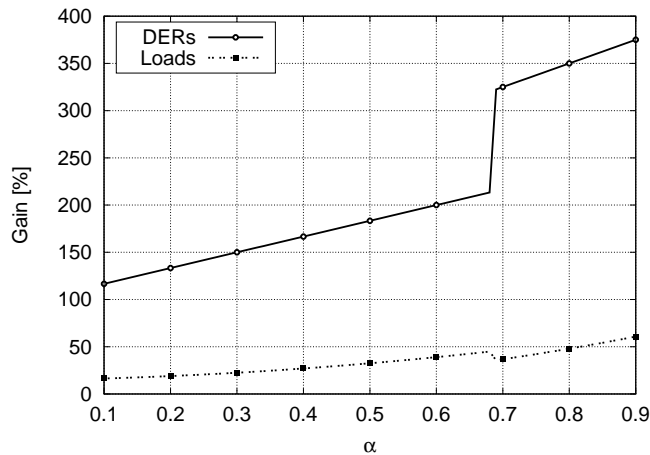}
\caption{Aggregated gains for revenues (DERs) and expenses (loads). Tight power offer case.
\label{figure:cap_5_tight_money}}
\end{figure}

\fig{figure:cap_5_tight_money} shows the DERs and the loads aggregate gains (with respect to the case in which no optimization is performed) obtained through the proposed optimization method, i.e., solving \eqref{eq:MOO_problem} for a discount factor $\alpha$ ranging from $10\%$ to $90\%$. We recall that $\alpha$ is used as a free parameter to bound the maximum expense from the PCC, according to the proposed discount strategy. A first noticeable result is that, for every value of $\alpha$, the optimized aggregate revenue is always larger than that in the non-optimized case. Moreover, the aggregate expense is always smaller than in the non-optimized one. These facts are highly desirable, since they guarantee that endorsing the proposed market model leads to a {\it substantial economic convenience} for all the agents involved in the energy trading process. 

\begin{figure}[t]
\centering
\includegraphics[scale=0.7]{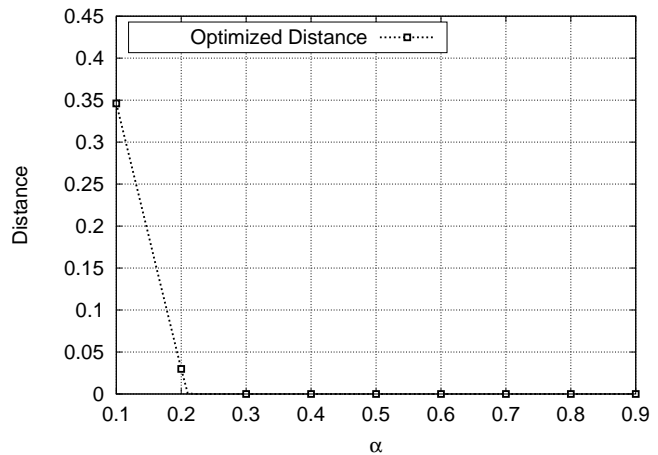}
\caption{Electrical efficiency in terms of distance from optimal electrical working point. Tight power offer case.
\label{figure:cap_5_tight_distance}}
\end{figure}

When computing the distance from the electrically efficient condition, the norm of of the difference $\Drow{i}-\DrowD{i}$ is computed for each load $i\in\mathcal{L}$. The plotted distance is thus the sum of the $L$ individual distances. \fig{figure:cap_5_tight_distance} shows this distance for $\alpha = 10\%,\dots,90\%$. We emphasize that, as the maximum discount factor reaches $21\%$, the electrical efficiency obtained through the proposed optimization equals the theoretical optimal electrical efficiency obtained through CBSC.

Remarkably, \fig{figure:cap_5_tight_money} and \fig{figure:cap_5_tight_distance} show that for a maximum discount factor of $21\%$, the \revision{P-}optimal solution reaches the maximum achievable electrical efficiency, while doubling the aggregate revenue of DERs with respect to the non-optimized case. At the same time, the consumers will incur sensibly smaller expenses.

\subsection{Scenario 2: Unbalanced Tight Power Offer}

In the unbalanced tight power offer scenario, $G_1$ is willing to sell more power than the amount dictated by CBSC, while $G_2$ sells less power than what dictated by CBSC.

\begin{figure}[t]
\centering
\includegraphics[scale=0.7]{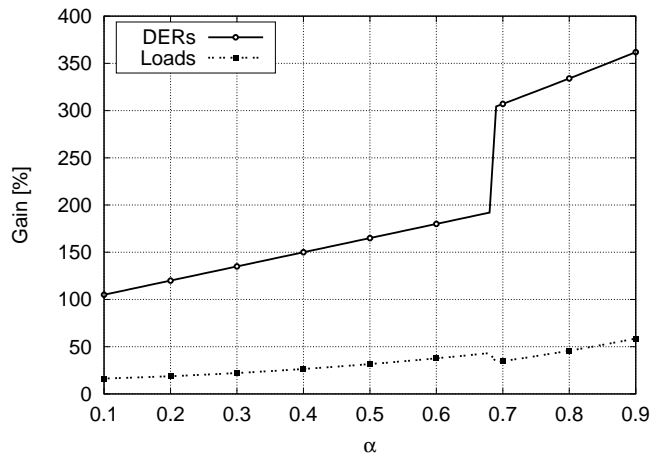}
\caption{Aggregated gains for revenues (DERs) and expenses (loads). Unbalanced tight power offer case.
\label{figure:cap_5_un_tight_money}}
\end{figure}

\fig{figure:cap_5_un_tight_money} shows the DERs and the loads aggregate gains obtained through the proposed optimization when the maximum discount factor $\alpha$ varies from $10\%$ to $90\%$. As in the previous case, endorsing the proposed optimization will lead to economical benefits for both the DERs and the loads.

\begin{figure}[t]
\centering
\includegraphics[scale=0.7]{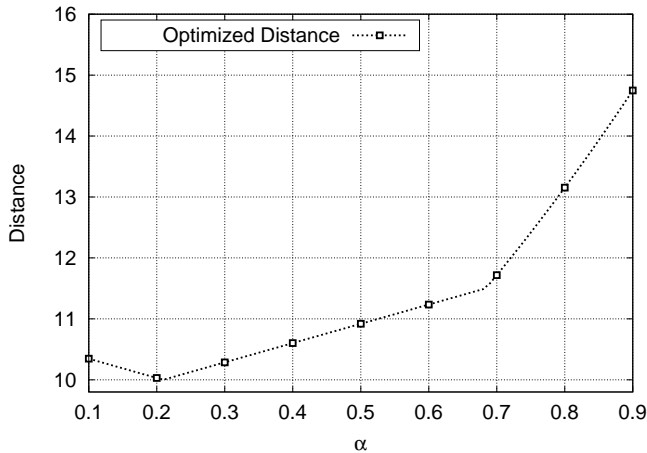}
\caption{Electrical efficiency in terms of distance from optimal electrical condition. Unbalanced tight power offer case.\label{figure:cap_5_un_tight_distance}}
\end{figure}

\fig{figure:cap_5_un_tight_distance} shows the distance between the power demand matrix obtained through the proposed optimization and the optimal one obtained through CBSC. The considered scenario does not allow to reach the theoretical optimal electrical efficiency. As a matter of fact, even though the total available power equals that required by CBSC, $G_1$ has more available power than what is needed, while $G_2$ has less. Hence, no configuration exists for which the power allocation matrix obtained through the proposed optimization approach can match the optimal power demand matrix. It can nevertheless be noted that, for a maximum discount factor of $\alpha=20\%$, the optimization process reaches the minimum achievable distance from the theoretical optimal working point. In contrast with the previous case, in this scenario there exists, for the selected weight vector $\bm{\lambda}$, a single maximum discount factor that allows to maximize the electrical grid efficiency (i.e., the one leading to the minimum in Fig.~\ref{figure:cap_5_un_tight_distance}). In fact, configurations exist where DERs and loads individual interests drive the grid toward a non-optimal power allocation condition, i.e., $G_1$, instead of selling $10$~kW to the PCC, it starts trading with $L_2$ leading to a sub-optimal electrical efficiency.

As for the previous case, \fig{figure:cap_5_un_tight_money} and \fig{figure:cap_5_un_tight_distance} show that the proposed optimization always ensures economical benefits for DERs and loads while, at the same time, leading to an increased electrical grid efficiency.

\subsection{Scenario 3: Loose Power Offer}

In the loose power offer scenario, $G_1$ sells more power than what dictated by CBSC, while $G_2$ sells the exact amount of power dictated by the CBSC algorithm.

\begin{figure}[t]
\centering
\includegraphics[scale=0.7]{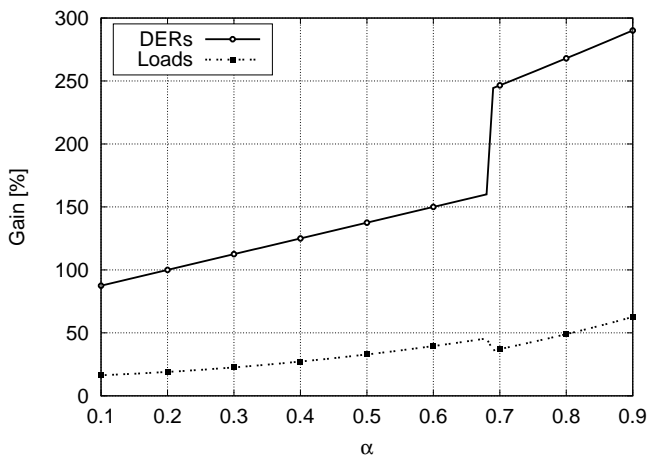}
\caption{Aggregated gains for revenues (DERs) and expenses (loads). Loose power offer case.
\label{figure:cap_5_loose_money}}
\end{figure}

\fig{figure:cap_5_loose_money} shows the performance of the proposed optimization in terms of the aggregated gains obtained by the DERs and the loads. As for the previous cases, we see that the proposed optimization always guarantees higher revenues and smaller expenses with respect to the case where the PCC is the only agent trading electrical power, i.e., all power has to be uniquely sold to or bough from the PCC.

\begin{figure}[t]
\centering
\includegraphics[scale=0.7]{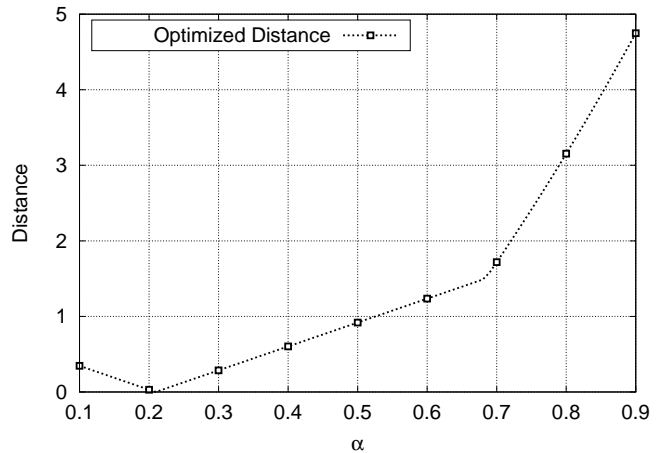}
\caption{Electrical efficiency in terms of distance from optimal electrical condition. Loose money case.\label{figure:cap_5_loose_distance}}
\end{figure}

\fig{figure:cap_5_loose_distance} shows that for $\alpha=20\%$ the optimal electrical working point is reached. In this case, $G_1$ sells $50$~kW to $L_1$ and the remaining available power is sold to the PCC. As $\alpha$ grows, $G_1$ starts selling more power to $L_1$ and hence the distance from the optimal electrical condition starts increasing. As for the previous case, for the selected weight vector $\bm{\lambda}$, a single value of $\alpha$ exists for which the electrical efficiency is maximized, i.e., the distribution power losses are minimized.

The presented results show that, for every considered power configuration, the proposed optimization approach results in substantial economical benefits \revision{(even though the objective functions of the DERs were given the smallest weight in the global optimization process)} and is likely to drive the power grid toward its maximum electrical efficiency. It is worth noting that, in the considered examples, the discount factor that is required to reach the electrical grid efficiency is never higher than $21\%$. This is appealing as it shows that the maximum discount remains rather small, irrespective of the network configuration. This may be especially convenient for the grid operator in practical scenarios.

%% file: conclusions.tex
%!TEX root = main.tex
In this paper, an original market model for smart grids was presented. The proposed framework jointly accounts for end users economical benefits and electrical grid efficiency maximization. This model was formally described as a non convex multi-objective optimization problem, which was then transformed into a convex one through a bijective transformation based on geometric programming. \revision{Pareto-}optimal trading and discount policies were devised through the solution of the equivalent convex formulation. \revision{Both a centralized and decentralized solution have been devised}. The performance of the proposed market model was then assessed in terms of electrical efficiency for the power grid and achievable economical benefit for all involved actors, i.e., profit made by DERs and expense incurred by the loads. 
Several network configurations were considered so as to systematically test the efficacy of the proposed market model.   
Numerical results show that considerable economical benefits can be reached for all agents and that the micro grid can be concurrently driven toward an optimal working point through the use of small discount factors from the regulating authority.

%Three power configurations were considered and the proposed optimization framework was evaluated for each of them for an example network setup. The first configuration relates to the case where the power availability of DERs matches the optimal power allocation matrix dictated by the CBSC algorithm. The second configuration relates to the case in which the total power availability from DERs exactly matches the total power demand from the loads, but the optimal configuration can not be reached as DERs are not able to individually inject the needed amount of power. The third configuration relates to the case where the power availability exceeds the total power demand. Numerical results showed that for each of the three considered configurations, the proposed market model guarantees considerable economical benefits for both the DERs and the loads. Moreover, it was shown that the smart grid can be always driven to a solution where its electrical efficiency is optimal and this entails the use of small discount factors.